\documentclass{article}
\usepackage{graphicx}
\usepackage{authblk}
\usepackage{amsmath}
\usepackage{graphicx}
\usepackage{cite}
\usepackage{subcaption}
\usepackage{float}
\usepackage{appendix}
\usepackage{bm}
\usepackage[colorlinks=true,linktocpage=true,
linkcolor=blue,citecolor=blue]{hyperref}
\usepackage[a4paper]{geometry}
\usepackage{amssymb}
\usepackage{hyperref}

\begin{document}
\large
\title{\bf{Investigating the Seebeck effect of the QGP medium using a novel relaxation time approximation model}}
\author[1]{Anowar Shaikh\thanks{anowar.19dr0016@ap.iitism.ac.in}}
\author[2]{Shubhalaxmi Rath\thanks{shubhalaxmi@academicos.uta.cl}}
\author[3]{Sadhana Dash\thanks{sadhana@phy.iitb.ac.in}}
\author[1]{Binata Panda\thanks{binata@iitism.ac.in}}
\affil[1]{Department of Physics\\ Indian Institute of Technology (Indian School of Mines) Dhanbad, Jharkhand 826004, India}
\affil[2]{Instituto de Alta Investigaci\'{o}n, Universidad de Tarapac\'{a}, Casilla 7D, Arica, Chile}
\affil[3]{Department of Physics\\ Indian Institute of Technology Bombay, Mumbai 400076, India}
\date{}
\maketitle
\begin{abstract}
The highly energetic particle medium formed in the ultrarelativistic heavy ion collision displays a notable difference in the temperatures between its central and peripheral regions. This temperature gradient can 
generate an electric field within the medium, a phenomenon referred to as the Seebeck effect. The magnitude of such electric field per unit temperature gradient in the limit of zero electric current is known as the Seebeck coefficient. We have estimated the Seebeck coefficient for a dense quark-gluon plasma medium by using the relativistic Boltzmann transport equation in the recently developed novel relaxation time approximation (RTA) model within the kinetic theory framework. This study explores the Seebeck coefficient of individual quark flavors as well as the entire partonic medium, with the emphasis on its dependence on the temperature and the chemical potential. Our observation indicates that, for given current quark masses, the magnitude of the Seebeck coefficient for each quark flavor as well as for the partonic medium decreases as the temperature rises and increases as the chemical potential increases. Furthermore, we have investigated the Seebeck effect by considering the partonic interactions described in perturbative thermal QCD within the quasiparticle model. In addition, we have presented a comparison between our findings and the results of the standard RTA model. We have observed that the Seebeck coefficient of the QGP medium gets conspicuously decreased in the novel RTA model as compared to that in the standard RTA model. A decreased Seebeck coefficient in the novel RTA model describes a smaller magnitude of induced electric field in the medium than that estimated by the standard RTA model. However, the rate of decline gets gradually smaller as the medium gets hotter for both the current quark mass scenario and the quasiparticle mass scenario. It is also found that, in the noninteracting 
case, the Seebeck coefficient possesses a slightly negative value in the high temperature region, unlike the quasiparticle description, where the Seebeck coefficient remains positive for the entire temperature range. 

\end{abstract}

\newpage

\section{Introduction}
The primary objective in the study of ultrarelativistic heavy ion collisions is to detect and study the existence of a new form of matter as predicted by the quantum chromodynamics (QCD). These
experiments at particle accelerators provide insight into the characteristics of fundamental particles like quarks and gluons. There is compelling evidence that the new form of matter, {\em i.e.} the quark-gluon plasma (QGP) matter has been detected at experimental facilities such as the Large Hadron Collider (LHC) \cite{Bzdak:2011yy, ALICE:2005vhb, ALICE:2004fvi} at CERN and the Relativistic Heavy Ion Collider (RHIC) \cite{Skokov:2009qp, BRAHMS:2004adc, STAR:2005gfr, PHENIX:2004vcz} at BNL. There also exist some observables which are regarded as the signatures of the presence of such a matter. These signatures include jet quenching \cite{Wang:1992qdg,Gyulassy:1999zd,PHENIX:2001hpc}, quarkonium suppression \cite{Blaizot:1996nq,Satz:2006kba,Rapp:2008tf}, elliptic flow \cite{Bhalerao:2006tp,Voloshin:2007pc}, photon and dilepton spectra \cite{Shuryak:1978ij, Kapusta:1991qp} etc. Since the discovery of QGP, numerous aspects of the strongly coupled QCD matter have been the focus of many investigations throughout the course of the last several decades. Quantitative data from the experiments on hadrons align with the theoretical results from ideal fluid dynamics, thus confirming the low viscosity of matter and its flow as a perfect fluid \cite{Kolb:2003dz,Romatschke:2007mq,Schenke:2010nt,Niemi:2012ry}. Since then, it has been a pertinent and 
challenging task to comprehend the strongly interacting medium produced at high temperatures through studying its transport coefficients. 

Different types of methodologies, such as the perturbative QCD and alternative effective models were used to determine the transport coefficients of quark matter \cite{Muronga:2006zx,Puglisi:2014sha,Mitra:2017sjo,Gupta:2003zh,Aarts:2014nba,Ding:2016hua,Nam:2012sg,Greif:2014oia,Feng:2017tsh,Buividovich:2010tn,Kurian:2017yxj,Hattori:2016cnt,Kurian:2018qwb,Rath:2019vvi,Rath:2021ryd,Shaikh:2022sky}. The strongly coupled fluid behavior of QGP has made possible the use of theoretical approaches intended for systems of this kind. In kinetic theory, a simplified structure for the collision integral is often used, which is referred to as the relaxation time approximation (RTA) \cite{Anderson:1974nyl}. The Bhatnagar-Gross-Krook (BGK) model \cite{Bhatnagar:1954zz} and the Welander model \cite{welander1954temperature} had previously accomplished progress in the nonrelativistic scenario. The relaxation time approximation has been used in several fields of physics. For the purpose of investigating the domain and microscopic basis of the relevance of hydrodynamics with relativistic effects, the relaxation time approximation has become an essential technique. The approximation given in ref. \cite{Anderson:1974nyl}, although widely used, has some fundamental shortcomings and violates some macroscopic and microscopic conservation laws. These fundamental challenges are resolved and the essential characteristics of the linearized Boltzmann collision operator are preserved by using a novel collision integral \cite{Rocha:2021zcw}. Using the novel collision integral, the transport coefficients associated with the heat and the charge in a QGP medium have been derived in recent works \cite{Rath:2024vyq, Shaikh:2024gsm}. 

In the present work, we explore the thermoelectric effect by calculating the Seebeck coefficient 
for the strongly interacting matter created in heavy ion collisions with a temperature 
gradient between the fireball's core and its peripheral region. In the presence of a temperature gradient, a finite gradient of charge carriers may be produced, which ultimately results in an electric field. When there is a difference in the temperature inside a conducting material, the charge carriers in the hotter zone move towards the cooler zone, producing an electric field. When the generated electric field 
is strong enough to inhibit further charge flow, the diffusion ends. The Seebeck coefficient measures the magnitude of the electric field per unit temperature gradient in a medium. This thermoelectric transport coefficient is calculated by setting the electric current to zero \cite{PhysRevB.68.125210, Callen1960}. Though there is a temperature gradient, the QGP medium should not be confused with a dynamical expanding QGP, rather we take it as a static homogeneous medium. A homogeneous system, by definition, lacks a spatial temperature gradient, since homogeneity denotes uniform environments. However, one can incorporate a temperature gradient manually as an external perturbation instead of an inherent characteristic of the medium. In this work, we assume the presence of a small temperature gradient in the medium, allowing the use of linear response theory to compute the transport coefficients. The Seebeck coefficient is obtained by considering the induced electric field due to a small manually imposed temperature gradient, rather than one that naturally develops from an expanding system. According to the `fireball analogy', the quark-gluon plasma produced in heavy ion collisions is not completely static, rather, it has a central hot zone surrounded by cooler peripheral regions. The temperature gradient in the present work replicates the configuration, with the center exhibiting higher temperature than the periphery, resulting in the charge separation and the Seebeck effect. This approach does not require complete hydrodynamic evolution of the medium. Unlike condensed matter systems, where only one kind of charge carrier participates in the transport process, in a medium with the strong interactions, both positive and negative charge carriers participate in the transport phenomena. The strongly interacting matter requires both a temperature gradient and finite baryon chemical potential in order to observe the thermoelectric effect. In the absence of the quark chemical potential, the numbers of particles and antiparticles are same, so no net thermoelectric effect is observed, thus resulting in no observable net Seebeck effect. The convention is to consider the sign of the Seebeck coefficient as positive when the thermoelectric current flows from the hotter end to the cooler end. Therefore, the sign of the Seebeck coefficient may be used to determine the sign of the dominating charge carriers in the system. It exhibits positive sign for the positive charge carriers and negative sign for the negative charge carriers. Recently, researchers have investigated the Seebeck effect in a hot hadron gas using the hadron resonance gas model \cite{Bhatt:2018ncr}. Additionally, the Seebeck effect in the QGP phase has been examined within the framework of the Nambu-Jona-Lasinio model in ref. \cite{Abhishek:2020wjm}. In the background of the magnetic field, the thermoelectric response of the hot QCD medium has been investigated in references \cite{Kurian:2021zyb,Zhang:2020efz,Dey:2020sbm,Dey:2021crc,Khan:2022apd,Arxiv:2405.12510}. Our main objective in the present work is to investigate the impact of the novel RTA collision integral on the Seebeck coefficient of the QGP medium. In our approach, we incorporate the medium effects through the quasiparticle model \cite{Bannur:2006ww}, 
where the medium dependence is observed through the dispersion relations of the thermally massive quarks and gluons. 

This work has been organized as follows. Section 2 describes the method for obtaining the Seebeck coefficient and addresses the relativistic Boltzmann transport equation (RBTE) in the novel relaxation time approximation model. Section 3 focuses on the determination of the Seebeck coefficient by solving the RBTE 
with the novel RTA collision integral. In section 4, we have discussed the results of the 
aforementioned thermoelectric coefficient by taking into account the current quark masses and the quasiparticle masses of the partons. Section 5 presents the summary of this work. 

\noindent \textbf{Notations and conventions:} The covariant derivative $\partial_\mu$ represents the derivative with respect to $x^\mu$, while $\partial^{(p)} _\mu$ represents the derivative with 
respect to $p^\mu$. The fluid four velocity  $u^\mu=(1,0,0,0)$ is normalized to unity in the local rest 
frame  $(u^\mu u_\mu=1)$. In this work, the subscript $f$ represents the flavor index, where $f$ takes the values $u$, $d$ and $s$. Furthermore, $q_f$, $g_f$, and $ \delta f_{f}$ ($\delta \bar {f}_{f}$) represent the electric charge, the degeneracy factor, and the infinitesimal change in the distribution function 
for the quark (antiquark) of the $f^{th}$ flavor, respectively. The symbol $m_f$ represents the current 
quark mass, with values of 3 MeV, 5 MeV, and 100 MeV for up, down, and strange quarks, respectively. In this work, $g_g=2(N^2_c-1)$ represents the gluonic degrees of freedom, and $g_f(g_{\bar{f}})=2N_c$ denotes the quark (antiquark) degrees of freedom (for each flavor of quark), where $N_c=3$ is the number of colors. 

\section{Relativistic Boltzmann transport equation in the context of the novel relaxation time approximation}
In this section, we provide a comprehensive framework to analyze the thermoelectric effect in a hot partonic medium. We use this framework to calculate the Seebeck coefficient for each individual species as well 
as for the composite medium. We begin with the Boltzmann transport equation for a single particle, {\em i.e.}, 
\begin{equation}\label{1} 	
\frac{\partial f_f}{\partial t}+ \frac{\bf{p}}{m}\cdot\nabla f_f + \bf{F}\cdot\frac{\partial \textit{f}_\textit{f}}{\partial \bf{p}}=\left(\frac{{\partial \textit{f}_\textit{f}}}{\partial \textit{t}}\right)_c 
,\end{equation}
where $f_{f}=f_{eq,f}+\delta f_f$, $\mathbf{F}$ represents the force field that affects the particles in the medium and $m$ represents the mass of the particle. The term on the right-hand side arises from particle collisions occurring inside the medium. Even for the most basic solution of eq. \eqref{1}, the complex nature of the collision term poses some difficulties. Due to the complexity of quantum scattering theory, we use the relaxation time approximation to address the collision term which applies when the deviation ($\delta f_f$) is much less than the equilibrium particle distribution ($f_{eq,f}$), {\em i.e.}, $\delta f_f \ll f_{eq,f}$. To allow the linearization of the relativistic Boltzmann transport equation, we may assume that the quark distribution function is close to equilibrium. 

The relativistic Boltzmann transport equation in its general relativistic covariant version for a QGP medium containing quarks, antiquarks and gluons is given by
\begin{equation}\label{R.B.T.E.}
	\begin{split}
		p^\mu \partial_\mu f_f+q_fF^{\mu \nu} p_\nu \partial^{(p)} _\mu f_f =\textbf{C}\bigl[\textit{f}_\textit{f}\hspace{1mm}\bigr]
	\end{split} 
.\end{equation}
Here, $\textbf{C}\bigl[\textit{f}_\textit{f}\hspace{1mm}\bigr]$ is the collision term and $F^{\mu \nu}$ represents the electromagnetic field strength tensor, whose components describe the electric and 
magnetic fields. We consider the components $F^{i0}=\textbf{E}$ and $F^{0i}=-\textbf{E}$ to see the effect of electric field. The relativistic Boltzmann transport equation can further be expressed as
\begin{equation}\label{R.B.T.E.(1)}
\textbf{p}\cdot\frac{\partial f_{eq,f}}{\partial \textbf{r}}+q_{f} \textbf{E}\cdot\textbf{p}\frac{\partial f_{eq,f}}{\partial p^0}+q_f p_0\textbf{E}\cdot\frac{\partial f_{eq,f}}{\partial \textbf{p}}=\textbf{C}\bigl[\textit{f}_\textit{f}\hspace{1mm}\bigr]
.\end{equation}
When using the relaxation time approximation (RTA), a relaxation term of the following form replaces the collision term: 
\begin{equation}\label{2}
\textbf{C}\bigl[\textit{f}_\textit{f}\hspace{1mm}\bigr]=-\frac{\omega_f}{\tau_f}\delta f_f
.\end{equation}
The relaxation time $\tau_{f}$ is the time needed to restore the disturbed system to its original equilibrium condition. A damping frequency is represented by the expression $\frac{1}{\tau_{f} }$ in this context. In addition to the adaptability of the model, there is no instantaneous conservation of the particle number and it appears to be a noticeable drawback. This model is not compatible with the principles of microscopic and macroscopic conservation laws. As a result, it presents some challenges when attempting to describe relativistic gases using the energy-dependent relaxation time or general matching conditions. To tackle this issue a novel relaxation time approximation can be used, where the collision term is replaced by a relaxation term whose form is given \cite{Rocha:2021zcw} by
\begin{multline}\label{3}	
	\textbf{C}\bigl[\textit{f}_\textit{f}\hspace{1mm}\bigr]=-\frac{\omega_f}{\tau_{fp}}\biggr[\delta f_f-
	\frac{\left\langle({\omega_f}/{\tau_{fp}})\delta f_f\right\rangle_0}{\left\langle{\omega_f}/{\tau_{fp}}\right\rangle_0}+
	P^{(0)}_1\frac{\left\langle({\omega_f}/{\tau_{fp}})P^{(0)}_1\delta f_f\right\rangle_0}{\left\langle({\omega_f}/{\tau_{fp}})P^{(0)}_1P^{(0)}_1\right\rangle_0}
	+ p^{\left\langle\mu\right\rangle}\frac{\left\langle({\omega_f}/{\tau_{fp}})p_{\left\langle\mu\right\rangle}\delta f_f\right\rangle_0}{(1/3)\left\langle({\omega_f}/{\tau_{fp}})p_{\left\langle\mu\right\rangle}p^{\left\langle\mu\right\rangle}\right\rangle_0}\Biggl].	
\end{multline}
Here, 
\begin{equation}\label{4}
	P^{(0)}_1=1-\frac{\left\langle{\omega_f}/{\tau_{fp}}\right\rangle_0}{\left\langle{\omega^2_f}/{\tau_{fp}}\right\rangle_0}\omega_f,
\end{equation}
\begin{equation}\label{5}
p^{\left\langle\mu\right\rangle}=\Delta^{\mu\nu}p_{\nu}	,
\end{equation}
\begin{equation}\label{6}
\Delta^{\mu\nu}=g^{\mu\nu}-u^{\mu}u^{\nu}
.\end{equation}
The momentum integrals are defined with respect to the local equilibrium distribution function $f_{eq,f}$ as follows, 
\begin{eqnarray}\label{7}
\nonumber\left\langle ...\right\rangle_0 &=& \int \frac{d^3p}{(2\pi)^3p_0} f_{eq,f}(...) \\ &=& \int dP f_{eq,f}(...)
,\end{eqnarray}
where $\tau_{fp}$ is the relaxation time and $\nu_f=\frac{1}{\tau_{fp}}$ denotes the collision frequency 
of the medium. The equilibrium distribution functions for $f^{th}$ flavor of quark and antiquark are respectively given by
\begin{eqnarray}
&&f_{eq,f}=\frac{1}{e^{\beta(\omega_f-\mu_f)}+1}, \\ 
&&\bar{f}_{eq,f}=\frac{1}{e^{\beta(\omega_f+\mu_f)}+1}
,\end{eqnarray}
where $\omega_f=\sqrt{\textbf{p}^2+m^2_f}$, $\beta=\frac{1}{T}$, and $\mu_f$ represents the chemical potential of $f^{th}$ flavor of quark. The momentum-dependent relaxation time is given \cite{Dusling:2009df,Dusling:2011fd,Kurkela:2017xis} by
\begin{equation}\label{p77}
\tau_{fp}(\omega_f)=(\beta \omega_f)^{\gamma} \tau_{f}
.\end{equation}
Here, $\gamma$ is an arbitrary constant which controls the energy dependence of the relaxation time and has different values for different theories, for example, in QCD kinetic theories, $\gamma=\frac{1}{2}$ \cite{Dusling:2009df} and in scalar field theories $\gamma=1$ \cite{Calzetta:PRD37'1988}. Many previously published studies on QGP transport characteristics had postulated a power-law relation 
between the relaxation time and the energy, with $\gamma=0.5$ \cite{Dusling:2009df,Dusling:2011fd,Kurkela:2017xis}. By including the scale dependency of the relaxation time ($\tau_{fp}$), we enhance the novel RTA approximation with features known in QCD. The choice of $\gamma$ guarantees that the transport coefficients such as electrical conductivity, shear viscosity and thermal conductivity align with anticipated behaviors in quark-gluon plasma. In the analogous QCD scenario, high energy particles mainly lose energy through inelastic gluon radiation. The relaxation time scales as $\tau_{f p}\propto E^{1/2}$ due to the Landau-Pomeranchuk-Migdal suppression effect. The energy loss rate of a high energy parton follows $\frac{dE}{dt} \sim E^{1/2}$, which means that it increases with energy, but at a sublinear rate. An increased value of $\gamma$ (e.g., $\gamma=1$) would result in an excessively fast increase of the relaxation time with energy, hence causing an overestimation of the transport coefficients. A smaller value of $\gamma$ (e.g., $\gamma=0$) might make it insufficiently reliant on energy, which contradicts the anticipated quark scattering behavior in QGP. Thus, we set the value of $\gamma$ to $0.5$ in the present work. In the above equation, $\tau_f$ is momentum-independent, whose expression for a high temperature QCD medium in the framework of relativistic kinetic theory is given \cite{Hosoya:1983xm} by
\begin{equation}\label{1e}
\tau_{f}=\frac{1}{5.1T\alpha^2_s \log(1/\alpha_s)\left[1+0.12(2N_f+1)\right]}
,\end{equation}
where $\alpha_s$ is the QCD running coupling constant, which is a function of both temperature and chemical potential, and it has the following \cite{kapusta1989finite} form, 
\begin{eqnarray}\label{ww2}
\alpha_s=\frac{g^2}{4\pi}=\frac{12\pi}{\left(11N_c-2N_f\right)\ln\left({\Lambda^2}/{\Lambda_{\rm\overline{MS}}^2}\right)}
~.\end{eqnarray}
Here, $\Lambda_{\rm\overline{MS}}=0.176$ GeV, $\Lambda=2\pi\sqrt{T^2+\mu_f^2/\pi^2}$ for electrically charged particles (quarks and antiquarks) and $\Lambda=2 \pi T$ for gluons. 

\section{Seebeck coefficient of the QGP medium}
In the presence of the thermal gradient, the charge carriers move from the higher temperature zone to the lower temperature zone. Consequently, an electric current is generated in the medium and the 
corresponding induced current density is expressed as
\begin{equation}\label{010}
\mathbf{J}=\sum_{f}  g_{f}\int \frac {d^3p}{(2\pi)^3} \frac{\mathbf{p}}{\omega_f}\bigl[q_{f} \delta f_f(x,p)+\bar{q}_{f} \delta \bar{f}_f(x,p)\bigr]
,\end{equation}
where $\mathbf{J}$ represents the spatial part of the current density vector. The symbols $\delta f_{f}$ and $\delta \bar {f}_{f}$ respectively represent the infinitesimal changes in the phase space distribution functions of the $f^{th}$ flavor of quark and antiquark. The relativistic Boltzmann transport equation (\ref{R.B.T.E.(1)}) can be expressed in terms of the collision integral in the novel relaxation time approximation as
\begin{multline}\label{10}
\textbf{p}\cdot\frac{\partial f_{eq,f}}{\partial \textbf{r}}+q_{f} \textbf{E}\cdot\textbf{p}\frac{\partial f_{eq,f}}{\partial p^0}+q_f p_0\textbf{E}\cdot\frac{\partial f_{eq,f}}{\partial \textbf{p}}=-\frac{\omega_f}{\tau_{fp}}\biggr[\delta f_f-
	\frac{\left\langle({\omega_f}/{\tau_{fp}})\delta f_f\right\rangle_0}{\left\langle{\omega_f}/{\tau_{fp}}\right\rangle_0}\\+
	P^{(0)}_1\frac{\left\langle({\omega_f}/{\tau_{fp}})P^{(0)}_1\delta f_f\right\rangle_0}{\left\langle({\omega_f}/{\tau_{fp}})P^{(0)}_1P^{(0)}_1\right\rangle_0}
	+ p^{\left\langle\mu\right\rangle}\frac{\left\langle({\omega_f}/{\tau_{fp}})p_{\left\langle\mu\right\rangle}\delta f_f\right\rangle_0}{(1/3)\left\langle({\omega_f}/{\tau_{fp}})p_{\left\langle\mu\right\rangle}p^{\left\langle\mu\right\rangle}\right\rangle_0}\Biggl]
.\end{multline}
The partial derivatives appearing in the above equation are calculated as follows, 
\begin{equation}\label{11}	
	\left.	\begin{aligned}
 \frac{\partial f_{eq,f}}{\partial \textbf{r}}=\beta^2 (\omega_f-\mu_f) f_{eq,f}(1-f_{eq,f}) \vec{\nabla} T(\textbf{r}) \\ 
           \frac{\partial f_{eq,f}}{\partial p^0}=-\beta f_{eq,f}(1-f_{eq,f}) \\ 
		\frac{\partial f_{eq,f}}{\partial \mathbf{p}}=-\beta \mathbf{v} f_{eq,f}(1-f_{eq,f}) 
	\end{aligned}
	\right\}
.\end{equation}
Using eq. (\ref{11}), the left-hand side of eq. (\ref{10}) can be written as
\begin{multline}\label{12}
\textbf{p}\cdot\frac{\partial f_{eq,f}}{\partial \textbf{r}}+q_{f} \textbf{E}\cdot\textbf{p}\frac{\partial f_{eq,f}}{\partial p^0}+q_f p_0\textbf{E}\cdot\frac{\partial f_{eq,f}}{\partial \textbf{p}}=\textbf{p} \beta^2 (\omega_f-\mu_f) f_{eq,f}(1-f_{eq,f}) \vec{\nabla} T(\textbf{r})\\-{2q_f\beta (\textbf{E}\cdot\textbf{p}) f_{eq,f}(1-f_{eq,f})}.
 \end{multline}
Based on the results of the calculations, the right-hand side of eq. (\ref{10}) can be simplified (for a detailed calculation see appendix \ref{appendix A}) into 
\begin{multline}\label{13}	
	-\frac{\omega_f}{\tau_{fp}}\biggr[\delta f_f-
	\frac{\left\langle({\omega_f}/{\tau_{fp}})\delta f_f\right\rangle_0}{\left\langle{\omega_f}/{\tau_{fp}}\right\rangle_0}+
	P^{(0)}_1\frac{\left\langle({\omega_f}/{\tau_{fp}})P^{(0)}_1\delta f_f\right\rangle_0}{\left\langle({\omega_f}/{\tau_{fp}})P^{(0)}_1P^{(0)}_1\right\rangle_0}
	+ p^{\left\langle\mu\right\rangle}\frac{\left\langle({\omega_f}/{\tau_{fp}})p_{\left\langle\mu\right\rangle}\delta f_f\right\rangle_0}{(1/3)\left\langle({\omega_f}/{\tau_{fp}})p_{\left\langle\mu\right\rangle}p^{\left\langle\mu\right\rangle}\right\rangle_0}\Biggl]=\\-\frac{\omega_f A}{\tau_{fp}} \delta f_f.
 \end{multline}
Now, with the help of eq. (\ref{12}) and  eq. (\ref{13}), one can solve eq. (\ref{10}) to get the expression of $\delta f_f$ as
\begin{equation}\label{14}
\delta f_f=-\frac{\textbf{p} \beta^2 \tau_{fp}}{\omega_fA}(\omega_f-\mu_f) f_{eq,f}(1-f_{eq,f}) \vec{\nabla} T(\textbf{r})+\frac{2q_f\beta\tau_{fp} (\textbf{E}\cdot\textbf{p}) f_{eq,f}(1-f_{eq,f})}{\omega_fA}
,\end{equation}
where 
\begin{multline}\label{15}
A=\left[1-\frac{\omega_f\int p^2 (f_{eq,f}/ \tau_{fp}) dp}{\int p^2\omega_f (f_{eq,f}/\tau_{fp})dp}\right]\frac{\int p^2 (f_{eq,f}/\tau_{fp})\Bigr[1-\Bigr(\frac{\int p^2 (f_{eq,f}/\tau_{fp})dp}{\int p^2 \omega_f (f_{eq,f}/\tau_{fp})dp}\Bigl)\omega_f\Bigl] dp}{\int p^2 (f_{eq,f}/\tau_{fp})\Bigr[1-\Bigr(\frac{\int p^2 (f_{eq,f}/\tau_{fp})dp}{\int p^2 \omega_f (f_{eq,f}/\tau_{fp})dp}\Bigl)\omega_f\Bigl]^2 dp}\\ +\frac{3p\int p^3 (f_{eq,f}/ \tau_{fp}) dp}{\int p^4 (f_{eq,f}/ \tau_{fp}) dp}
.\end{multline}
Similarly for antiquarks, we have 
\begin{equation}\label{16}
\delta {\bar{f}}_f=-\frac{\textbf{p} \beta^2 \tau_ {\bar{fp}}}{\omega_f\bar{A}}(\omega_f+\mu_f){\bar{f}}_{eq,f} (1- {\bar{f}}_{eq,f}) \vec{\nabla} T(\textbf{r})+\frac{2q_ {\bar{f}}\beta\tau_ {\bar{fp}} (\textbf{E}\cdot\textbf{p}) {\bar{f}}_{eq,f}(1- {\bar{f}}_{eq,f})}{\omega_f\bar{A}},
\end{equation}
where 
\begin{multline}\label{17}
\bar{A}=	\left[1-\frac{\omega_f\int p^2 ({\bar{f}}_{eq,f}/ \tau_ {\bar{fp}}) dp}{\int p^2\omega_f ({\bar{f}}_{eq,f}/\tau_ {\bar{fp}})dp}\right]\frac{\int p^2 ({\bar{f}}_{eq,f}/\tau_ {\bar{fp}})\Bigr[1-\Bigr(\frac{\int p^2 ({\bar{f}}_{eq,f}/\tau_ {\bar{fp}})dp}{\int p^2 \omega_f ({\bar{f}}_{eq,f}/\tau_ {\bar{fp}})dp}\Bigl)\omega_f\Bigl] dp}{\int p^2 ({\bar{f}}_{eq,f}/\tau_ {\bar{fp}})\Bigr[1-\Bigr(\frac{\int p^2 ({\bar{f}}_{eq,f}/\tau_ {\bar{fp}})dp}{\int p^2 \omega_f ({\bar{f}}_{eq,f}/\tau_ {\bar{fp}})dp}\Bigl)\omega_f\Bigl]^2 dp}\\ 
+\frac{3p\int p^3 ({\bar{f}}_{eq,f}/ \tau_ {\bar{fp}}) dp}{\int p^4 ({\bar{f}}_{eq,f}/ \tau_ {\bar{fp}}) dp}
.\end{multline}
In order to get the spatial component of the induced current density, let us replace the values of $\delta f_f$ and $\delta {\bar{f}}_f$ in eq. (\ref{010}) for a single quark flavor as
\begin{multline}\label{18}
	\textbf{J}= \frac{g_{f}q_{f}}{2\pi^2} \Biggl[\int dp \frac{p^4}{\omega^2_f}\Biggl[\frac{\tau_{fp}f_{eq,f}(1-f_{eq,f})(\omega_f-\mu_f)}{A}-\frac{\tau_ {\bar{fp}} {\bar{f}}_{eq,f}(1-{\bar{f}}_{eq,f})(\omega_f+\mu_f)}{\bar{A}}
	 \Biggr]\Bigl(-\beta^2 \vec{\nabla}T(\textbf{r})\Bigr)\\+ 2 q_{f}\int dp \frac{p^4}{\omega^2_f}\Biggl[\frac{\tau_{fp}f_{eq,f}(1-f_{eq,f})}{A}+\frac{\tau_ {\bar{fp}} {\bar{f}}_{eq,f}(1-{\bar{f}}_{eq,f})}{\bar{A}}
	 \Biggr]\Bigl({\beta}\textbf{E}\Bigr)\Biggr].
\end{multline}
When the system is in a condition of equilibrium, the net current resulting from the $f^{th}$ quark flavor becomes zero. The current density mentioned above (eq. \eqref{18}) is assumed to be zero in order to 
determine the electric field resulting from the temperature gradient in the medium. This allows us to establish a relationship between the temperature gradient in the coordinate space and the induced electric field inside the medium as
\begin{equation}\label{19}
\textbf{E}=S \vec{\nabla} T(\textbf{r}).
\end{equation}
Here, the Seebeck coefficient ($S$) is defined as the electric field (magnitude) created in a conducting medium per unit temperature gradient when the electric current is set to zero and thus, we have 
 \begin{equation}\label{20}
     S=\frac{\beta}{2q_f}\frac{\int dp \frac{p^4}{\omega^2_f}\Bigl[\frac{\tau_{fp}f_{eq,f}(1-f_{eq,f})(\omega_f-\mu_f)}{A}-\frac{\tau_ {\bar{fp}} {\bar{f}}_{eq,f}(1-{\bar{f}}_{eq,f})(\omega_f+\mu_f)}{\bar{A}}
	 \Bigr]}{\int dp \frac{p^4}{\omega^2_f}\Bigl[\frac{\tau_{fp}f_{eq,f}(1-f_{eq,f})}{A}+\frac{\tau_ {\bar{fp}} {\bar{f}}_{eq,f}(1-{\bar{f}}_{eq,f})}{\bar{A}}
	 \Bigr]}.
 \end{equation}
For a single species and degeneracy factor, eq. (\ref{20}) may be recast into 
\begin{equation}\label{21}
S_f=\frac{\beta}{2q_f}\frac{Q_2}{Q_1}
,\end{equation}
where, 
\begin{equation}\label{22}
Q_2=\int dp ~ \frac{p^4}{\omega^2_f}\Bigl[\frac{\tau_{fp}f_{eq,f}(1-f_{eq,f})(\omega_f-\mu_f)}{A}-\frac{\tau_ {\bar{fp}} {\bar{f}}_{eq,f}(1-{\bar{f}}_{eq,f})(\omega_f+\mu_f)}{\bar{A}}
\Bigr]
,\end{equation}
\begin{equation}\label{23}
Q_1=\int dp ~ \frac{p^4}{\omega^2_f}\Bigl[\frac{\tau_{fp}f_{eq,f}(1-f_{eq,f})}{A}+\frac{\tau_ {\bar{fp}} {\bar{f}}_{eq,f}(1-{\bar{f}}_{eq,f})}{\bar{A}}\Bigr]
.\end{equation}
The analogous expressions for $Q_2$ and $Q_1$ in the standard RTA model are given \cite{Dey:2020sbm} by
\begin{equation}\label{Q2.R.T.A.}
Q_2=\int dp ~ \frac{p^4}{\omega^2_f}\Bigl[{f_{eq,f}(1-f_{eq,f})(\omega_f-\mu_f)}- { {\bar{f}}_{eq,f}(1-{\bar{f}}_{eq,f})(\omega_f+\mu_f)}
\Bigr]
,\end{equation}
\begin{equation}\label{Q1.R.T.A.}
Q_1=\int dp ~ \frac{p^4}{\omega^2_f}\Bigl[{f_{eq,f}(1-f_{eq,f})}+{ {\bar{f}}_{eq,f}(1-{\bar{f}}_{eq,f})}\Bigr]
.\end{equation}
Thus far, our analysis has included a single quark flavor. However, we will use the QGP medium that comprises of $u$, $d$ and $s$ quark flavors. 

The overall spatial current in the medium is the resultant vector of the currents caused by the individual species and is written as
\begin{eqnarray}\label{24}
\nonumber\textbf{J} &=& \sum_{f} J^f
 = J^{u} +J^{d}+J^{s} \\ \nonumber &=& \left[\frac{g_{u}q^2_{u}\beta}{ \pi^2} (Q_1)_u+\frac{g_{d}q^2_{d}\beta}{ \pi^2} (Q_1)_d+\frac{g_{s}q^2_{s}\beta}{ \pi^2} (Q_1)_s\right]\textbf{E} \\ && - \left[\frac{g_{u}q_{u}\beta^2}{2 \pi^2} (Q_2)_u+\frac{g_{d}q_{d}\beta^2}{2 \pi^2} (Q_2)_d+\frac{g_{s}q_{s}\beta^2}{2 \pi^2} (Q_2)_s\right]\vec{\nabla} T(\textbf{r})
.\end{eqnarray}
The total induced electric field at steady state is obtained by setting the current density \textbf{J} 
equal to zero, 
\begin{equation}\label{25}
\textbf{E}=\frac{\sum_{f} g_{f}q_{f} (Q_2)_f}{\sum_{f} g_{f}q^2_{f} (Q_1)_f} \frac{\beta}{2} \vec{\nabla} T(\textbf{r})
,\end{equation}
which provides the total Seebeck coefficient for the whole medium as
\begin{equation}\label{26}
S=\frac{\beta}{2} \frac{\sum_{f} g_{f}q_{f} (Q_2)_f}{\sum_{f} g_{f}q^2_{f} (Q_1)_f}.
\end{equation}
The overall Seebeck coefficient of the medium can be expressed in terms of the Seebeck coefficient of each flavor (since each flavor has the same degeneracy factor ($g_f$)) as
\begin{equation}\label{27}
S=\frac{\sum_{f} q^2_{f} S_f (Q_1)_f}{\sum_{f} q^2_{f} (Q_1)_f}
.\end{equation}

\section{Results and discussions}
The quasiparticle description provides a phenomenological explanation of the behavior of quarks and gluons in a thermal QCD medium. This description takes into account the generation of thermal masses for the partons, 
in addition to their current masses. Thermal masses get generated when the partons interact with the medium and one may perceive QGP as a collection of the noninteracting quasiparticles inside the framework 
of the quasiparticle model. A model of this kind was first introduced by Goloviznin and Satz \cite{Goloviznin:1992ws} to investigate the properties of the gluonic plasma. Subsequently, Peshier {\em et al.} \cite{Peshier:1995ty, Peshier:2002ww} used this model to examine the equation of state of the quark-gluon plasma derived from the lattice QCD calculations at finite temperature. In references \cite{Bluhm:2004xn, Bluhm:2007nu}, the quasiparticle model has been used to elucidate the lattice data, where a proper description of quasiparticles for the QGP medium has been utilized by taking into account the temperature and chemical potential dependent 
masses. The findings obtained from these models indicate that it is feasible to 
characterize the high temperature QGP phase by using a thermodynamically coordinated quasiparticle model. 
In this work, we have used the quasiparticle model, where the effective mass of the $f^{th}$ quark flavor is written \cite{Bannur:2006ww,Bannur:2005wm} as
\begin{equation}\label{28}
m_{fT}^2=m^2_f+\sqrt{2}m_f m_{ft}+ m^2_{ft}
,\end{equation}
where $m_f$ and $m_{ft}$ represent the current mass and the thermal mass of the quark, respectively. The latter can be computed from high temperature perturbative QCD approach using the hard thermal loop (HTL) approximation, where higher order coupling terms have been truncated and is given \cite{Peshier:2002ww,Braaten:1991gm,Bellac:BOOK'1996} by
\begin{equation}\label{29}
m_{ft}^2=\frac{g^2T^2}{6}\left(1+\frac{\mu_f^2}{\pi^2T^2}\right)
.\end{equation}
Here, $g$ is computed within the perturbative domain \cite{kapusta1989finite,Hosoya:1983xm} and is given by the equation $g=(4\pi\alpha_s)^{{1}/{2}}$, where $\alpha_s$ is defined in equation \eqref{ww2}. From eq. \eqref{ww2}, it is evident that the value of $\alpha_s$ is affected by both temperature and chemical potential. For a pure thermal QCD medium, the available energy scales are associated with the temperature, the chemical potential and the quark mass. In the QGP phase consisting of light quarks, antiquarks and gluons, the temperature is strong enough to become the largest energy scale. In this strong temperature regime, the divergences encountered in the calculation of QCD thermodynamic observables, transport coefficients and amplitudes are cured by applying the effective theories, one of which is the hard thermal loop perturbation theory. In this theory, the loop momentum is of the order of $T$, {\em i.e.} the hard scale and the next scale is the soft scale of the order of $gT$. The HTL perturbation theory could be used without contravening the definitions of thermodynamic quantities and transport coefficients. In our recent work \cite{Shaikh:2022sky}, we compared the findings of electrical conductivity with the specified coupling constant with the results of lattice QCD (lQCD) simulations \cite{Aarts:2014nba,Amato:PRL111'2013,Ding:LATTICE2014'2015}. We found that the values of electrical conductivity are comparable to lattice findings at high temperatures, however, the discrepancy is significant at low temperatures. While the aforementioned coupling constant accurately characterizes the asymptotic high temperature behavior of QCD, it exhibits some limitations when compared with lQCD data. Leading order pertubative QCD (pQCD) suggests that the system behaves as a weakly coupled environment, keeping higher order quantum corrections negligible. Nevertheless, lattice QCD simulations indicate that the QGP continues to exhibit strong interactions near the critical temperature ($T_c$), necessitating a greater effective coupling than that forecasted by leading-order pQCD calculations \cite{Plumari:PRD84'2011,Song:PRC93'2016,Sambataro:EPJC84'2024,Soloveva:PRC110'2024}. In the quasiparticle model, the dispersion relation for the $f^{th}$ flavor of quark is expressed as
\begin{equation}\label{30}
\omega^2_f=p^2+m^2_f+\sqrt{2}m_f m_{ft}+ m^2_{ft}
.\end{equation}

\begin{figure}[H]
\begin{subfigure}[h]{0.5\textwidth}
\includegraphics[width=\textwidth]{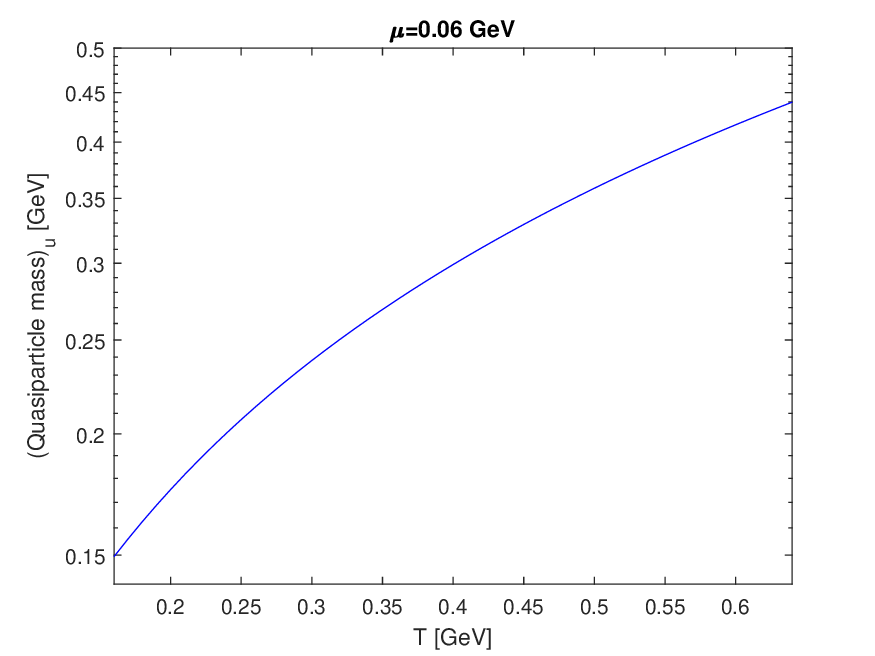}
\caption{}\label{uf1}
\end{subfigure}
\hfill
\begin{subfigure}[h]{0.5\textwidth}
\includegraphics[width=\textwidth]{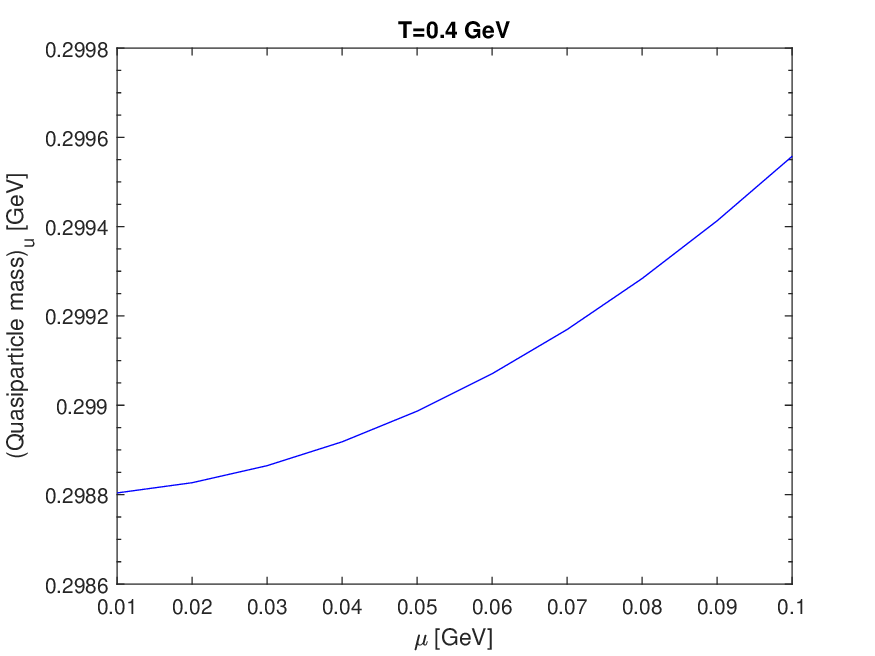}
\caption{}\label{uf2}
\end{subfigure}
\caption{The variation of the quasiparticle mass for $u$ quark (a) with temperature at a fixed chemical potential and (b) with chemical potential at a fixed temperature.}\label{u}
\end{figure}

\begin{figure}[H]
\begin{subfigure}[h]{0.5\textwidth}
\includegraphics[width=\textwidth]{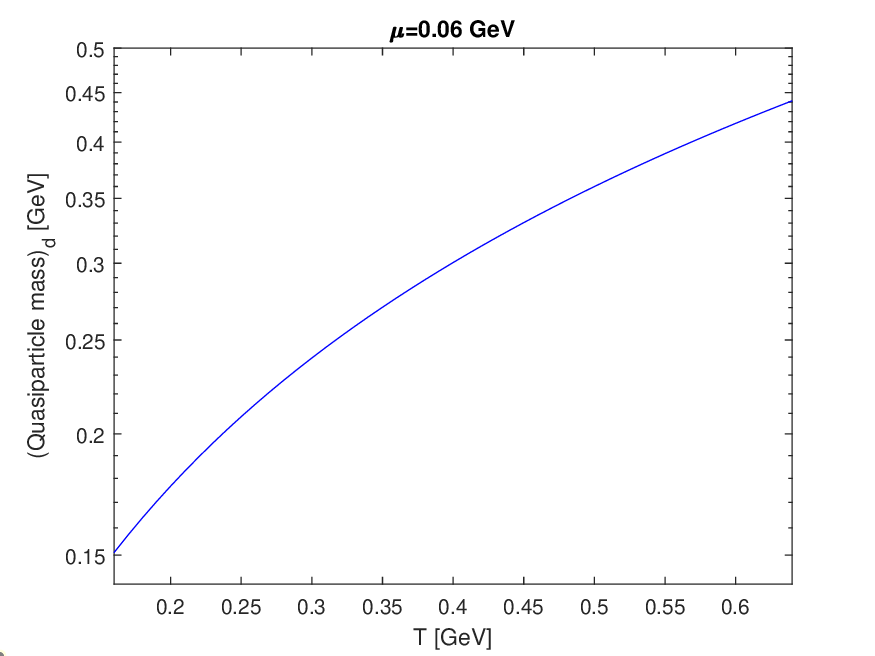}
\caption{}\label{df1}
\end{subfigure}
\hfill
\begin{subfigure}[h]{0.5\textwidth}
\includegraphics[width=\textwidth]{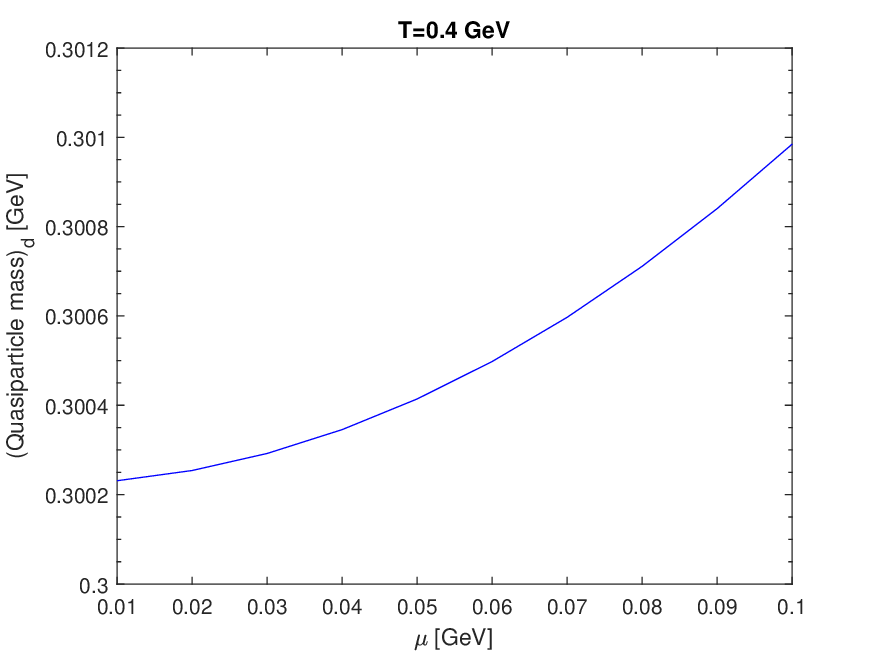}
\caption{}\label{df2}
\end{subfigure}
\caption{The variation of the quasiparticle mass for $d$ quark (a) with temperature at a fixed chemical potential and (b) with chemical potential at a fixed temperature.}\label{d}
\end{figure}

\begin{figure}[H]
\begin{subfigure}[h]{0.5\textwidth}
\includegraphics[width=\textwidth]{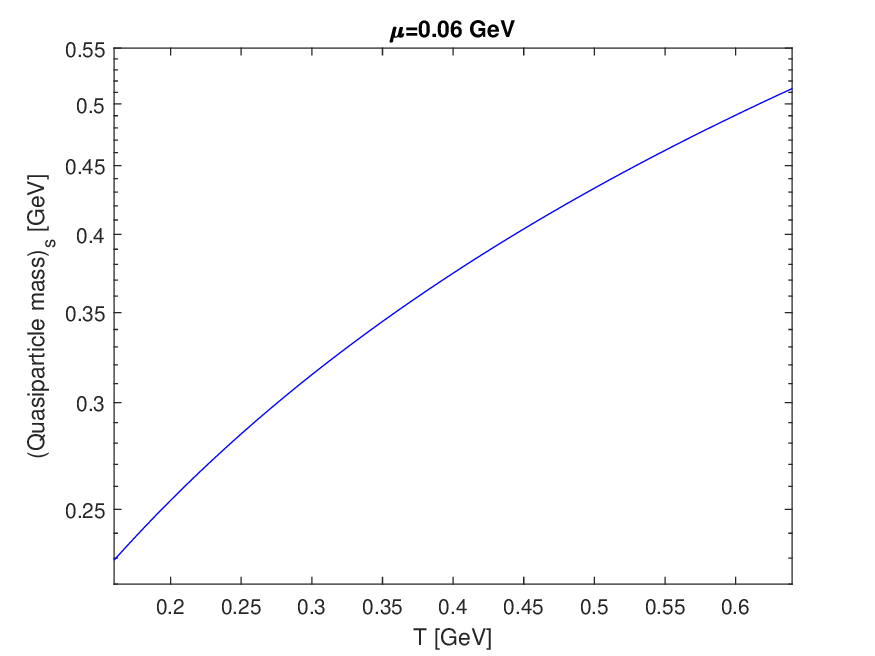}
\caption{}\label{sf1}
\end{subfigure}
\hfill
\begin{subfigure}[h]{0.5\textwidth}
\includegraphics[width=\textwidth]{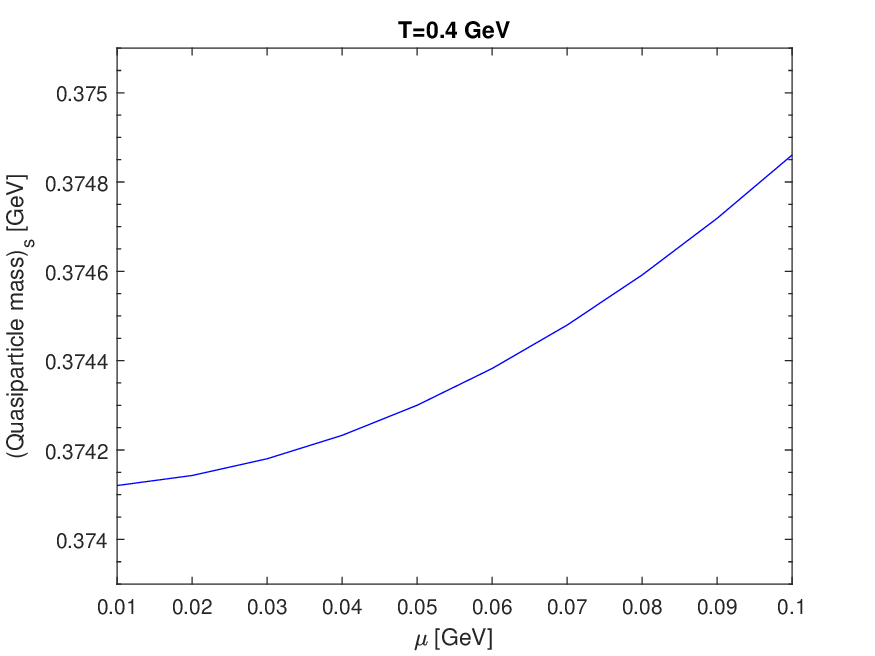}
\caption{}\label{sf2}
\end{subfigure}
\caption{The variation of the quasiparticle mass for $s$ quark (a) with temperature at a fixed chemical potential and (b) with chemical potential at a fixed temperature.}\label{s}
\end{figure}

In figures \ref{u}, \ref{d} and \ref{s}, we have plotted the quasiparticle masses of $u$, $d$ and $s$ quarks, respectively. With the increase of the temperature, quasiparticle masses are seen to increase (figures \ref{uf1}, \ref{df1} and \ref{sf1}). Similar increasing behaviors of quasiparticle masses are observed when we increase the chemical potential (figures \ref{uf2}, \ref{df2} and \ref{sf2}). Thus, there is an overall increase in the quark masses when shifting from current mass case to quasiparticle mass case. This significant change in the quark masses can contribute to major difference between the Seebeck coefficients calculated 
in the current quark mass scenario and the quasiparticle mass scenario. 

\begin{figure}[H]
\begin{subfigure}[h]{0.5\textwidth}
\includegraphics[width=\textwidth]{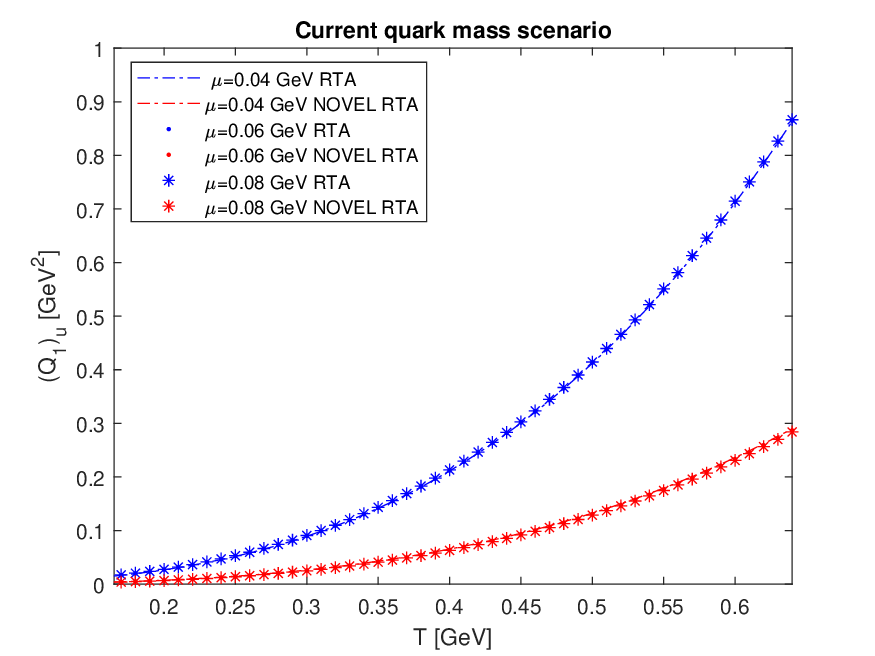}
\caption{}\label{1f1}
\end{subfigure}
\hfill
\begin{subfigure}[h]{0.5\textwidth}
\includegraphics[width=\textwidth]{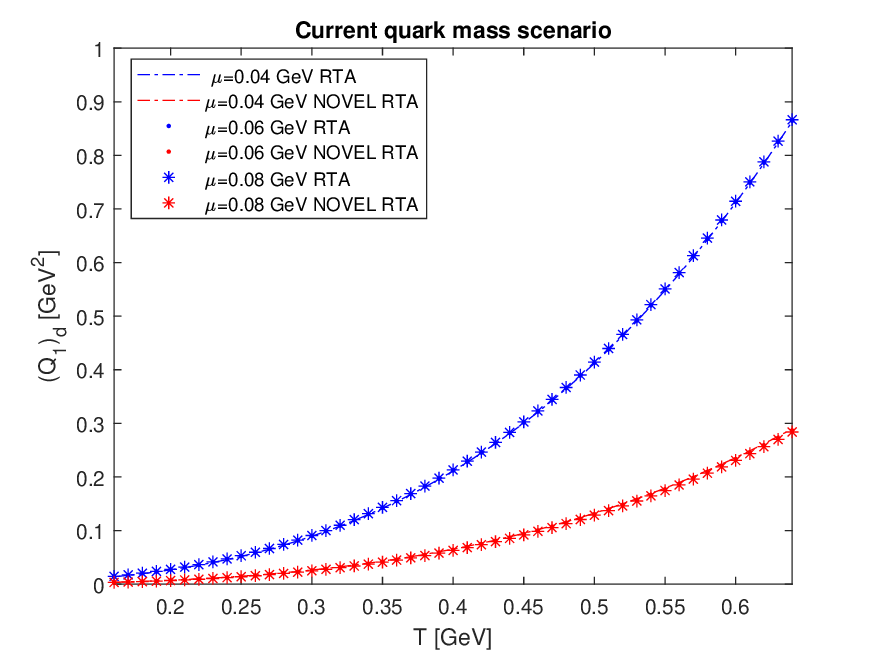}
\caption{}\label{1f2}
\end{subfigure}
\caption{The variation of $Q_1$ as a function of temperature (a) for $u$ quark and (b) for $d$ quark.}
\end{figure}

\begin{figure}[H]
\begin{subfigure}[h]{0.5\textwidth}
\includegraphics[width=\textwidth]{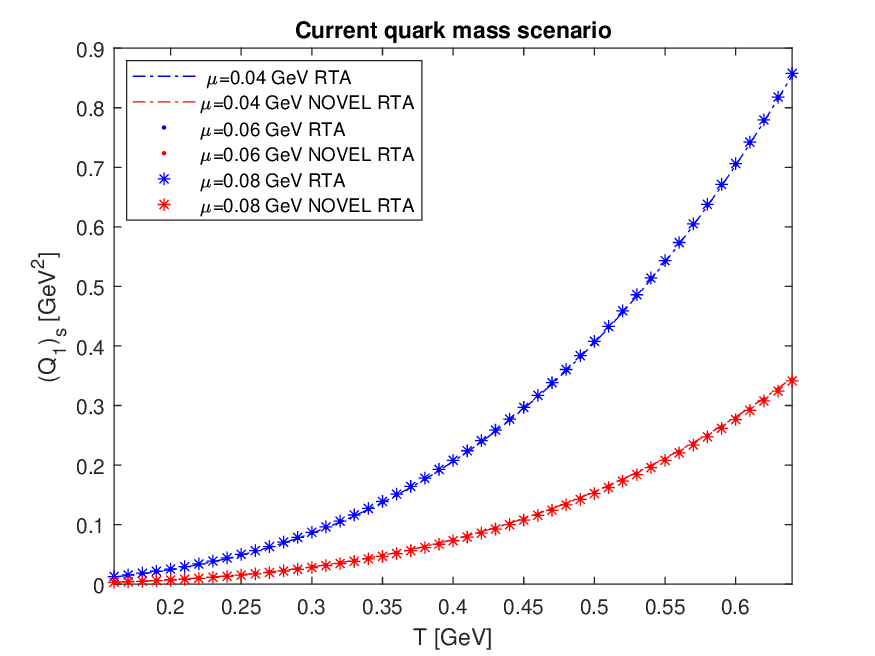}
\caption{}\label{2f1}
\end{subfigure}
\hfill
\begin{subfigure}[h]{0.5\textwidth}
\includegraphics[width=\textwidth]{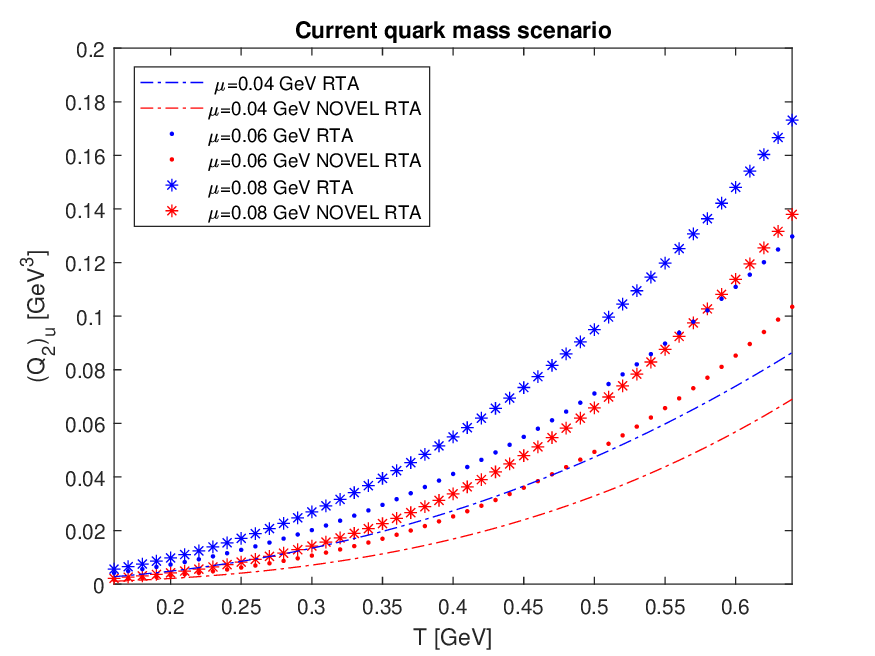}
\caption{}\label{2f2}
\end{subfigure}
\caption{(a) The variation of $Q_1$ as a function of temperature for $s$ quark and (b) the variation of $Q_2$ as a function of temperature for $u$ quark.}
\end{figure}

\begin{figure}[H]
\begin{subfigure}[h]{0.5\textwidth}
\includegraphics[width=\textwidth]{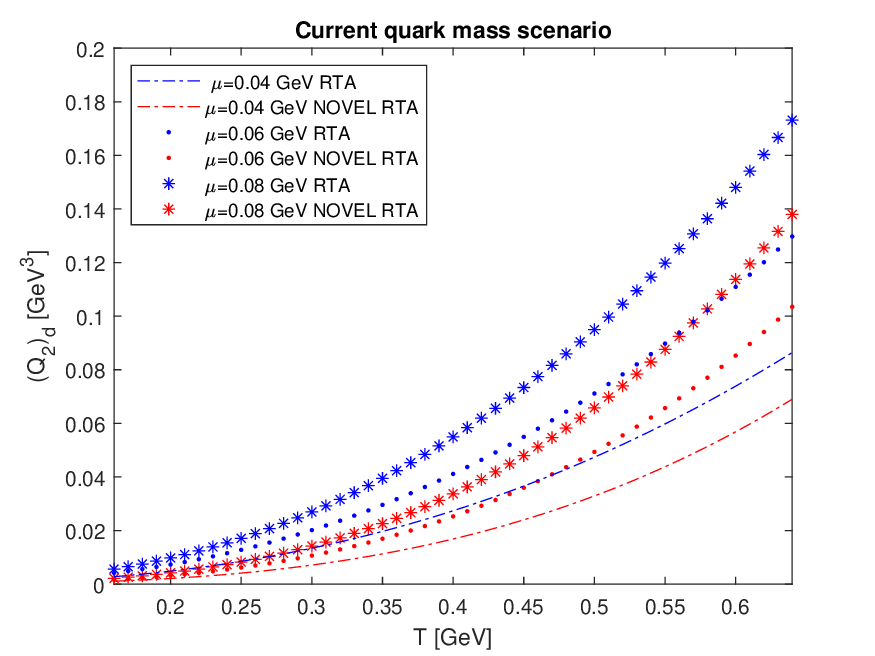}
\caption{}\label{3f1}
\end{subfigure}
\hfill
\begin{subfigure}[h]{0.5\textwidth}
\includegraphics[width=\textwidth]{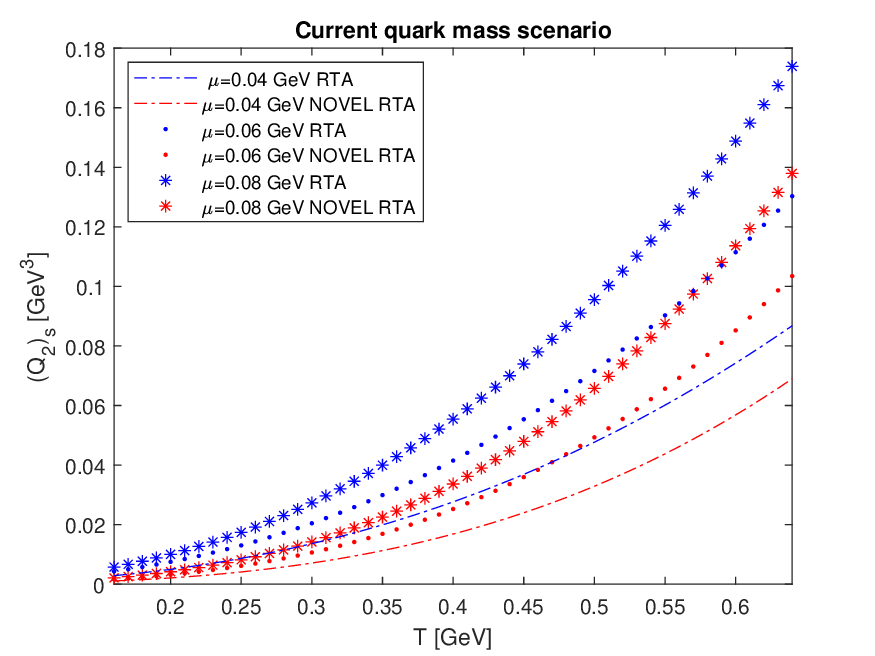}
\caption{}\label{3f2}
\end{subfigure}
\caption{The variation of $Q_2$ as a function of temperature (a) for $d$ quark and (b) for $s$ quark.}
\end{figure}

Using equations \eqref{21}, \eqref{22}, \eqref{23} and \eqref{27}, we have calculated the Seebeck coefficient for each species as well as for the medium for quark chemical potential $\mu=0.04$ GeV, $0.06$ GeV, $0.08$ GeV. We have used a temperature range from 0.16 GeV to 0.64 GeV for different chemical potential values. The variations of the integrals $Q_1$ and $Q_2$ with the temperature for $u$, $d$ and $s$ quarks at different finite chemical potentials are shown in figures \ref{1f1}-\ref{2f1} and \ref{2f2}-\ref{3f2}, respectively 
for the current quark mass scenario. The figures also compare our findings for the integrals $Q_1$ and $Q_2$ with the results obtained using the standard RTA model. It has been found that the modified RTA collision term leads to a decrease in these integrals for the QGP medium, regardless of the quark flavor. The closely comparable current masses of the $u$ and $d$ quarks result in similar values of the $Q_1$ and $Q_2$ integrals, whereas the values of these integrals are slightly different for the $s$ quark (as seen in figures \ref{2f1} and \ref{3f2}) due of its large mass. However, the ratio $Q_2/Q_1$ for $s$ quark produces values that are not significantly different from those of the $u$ and $d$ quarks. This observation on the 
integrals $Q_1$ and $Q_2$ facilitates the understanding of the Seebeck coefficient for individual quarks. 

\begin{figure}[H]
\begin{subfigure}[h]{0.5\textwidth}
\includegraphics[width=\textwidth]{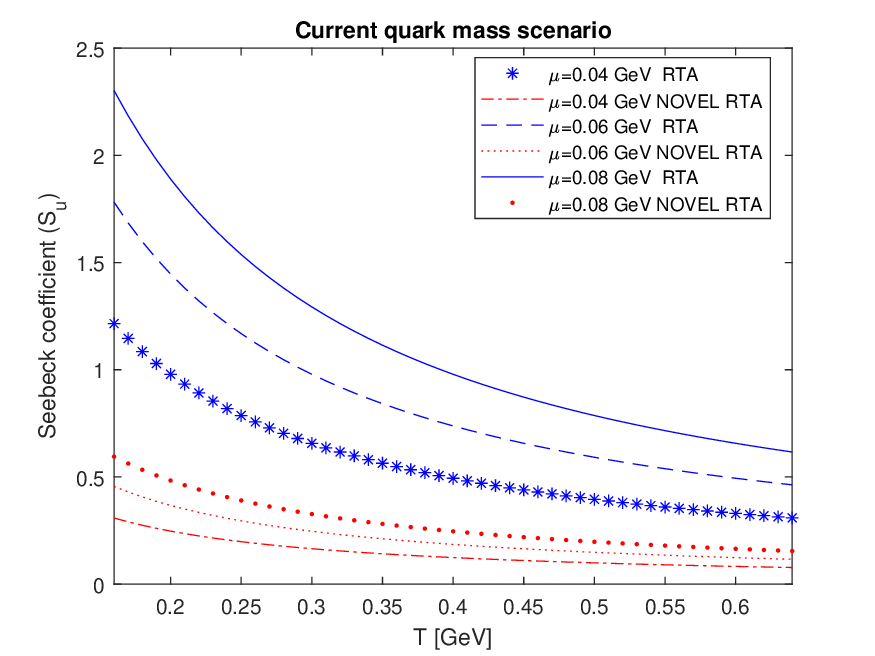}
\caption{}\label{4f1}
\end{subfigure}
\hfill
\begin{subfigure}[h]{0.5\textwidth}
\includegraphics[width=\textwidth]{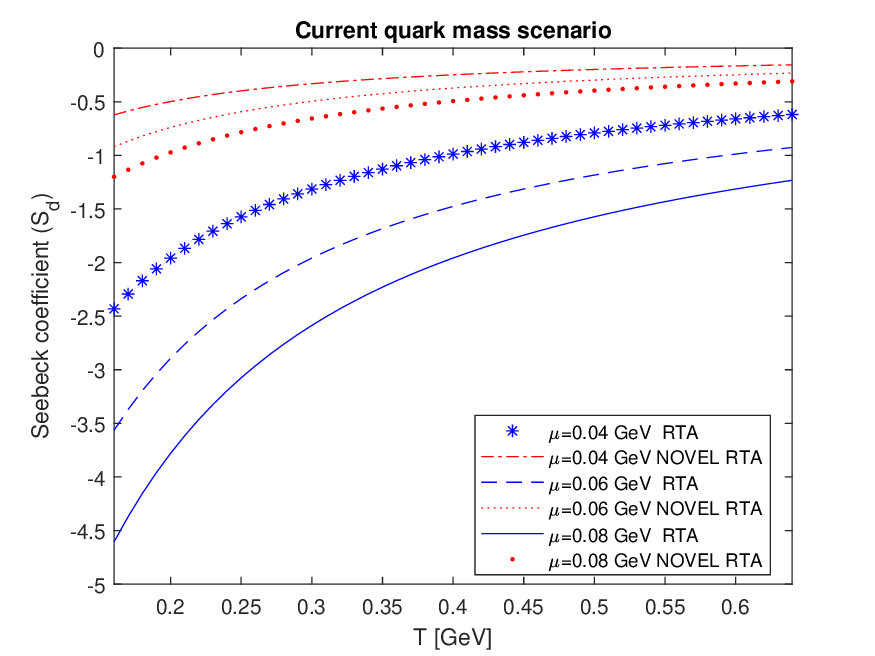}
\caption{}\label{4f2}
\end{subfigure}
\caption{The variation of Seebeck coefficient as a function of temperature (a) for $u$ quark and (b) for $d$ quark.}
\end{figure}

\begin{figure}[H]
\begin{subfigure}[h]{0.5\textwidth}
\includegraphics[width=\textwidth]{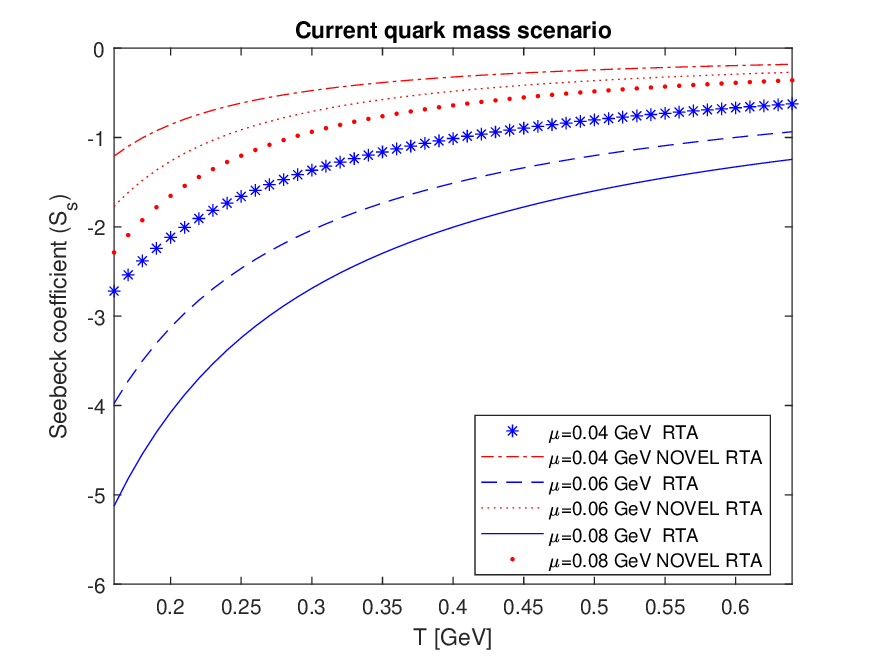}
\caption{}\label{5f1}
\end{subfigure}
\hfill
\begin{subfigure}[h]{0.5\textwidth}
\includegraphics[width=\textwidth]{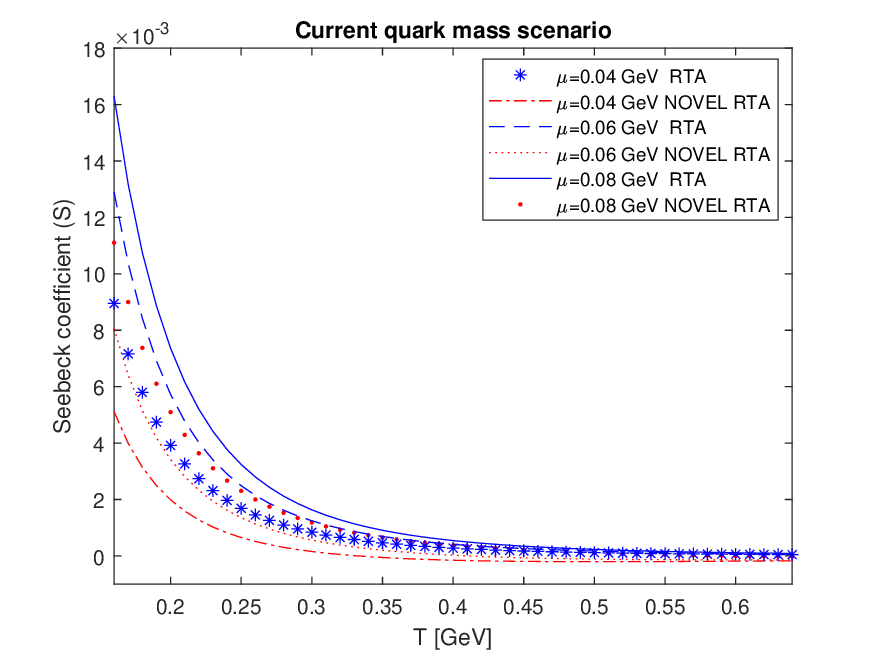}
\caption{}\label{5f2}
\end{subfigure}
\caption{(a) The variation of Seebeck coefficient as a function of temperature for $s$ quark and (b) the variation of the total Seebeck coefficient as a function of temperature.}
\end{figure}

Figures \ref{4f1}-\ref{5f1} and \ref{5f2} demonstrate the changes in the Seebeck coefficient with an increase 
of the temperature for individual quarks and the composite system at various finite chemical potentials, respectively. Based on the information shown in figures \ref{4f1}-\ref{5f1}, it is observed that the magnitude 
of the Seebeck coefficient for each quark decreases as the temperature increases, while keeping the chemical potential constant, for both collision terms. The reduction in the overall particle number density as the temperature increases, while the chemical potential remains constant, accounts for the decreasing value in the Seebeck coefficient. On the other hand, it has been observed that the Seebeck coefficient rises with the increasing chemical potential at a certain temperature. This is due to the fact that a higher chemical potential is suggestive of a larger number of particles over antiparticles. The Seebeck coefficients for $d$ and $s$ quarks undergo a reversal in sign as a result of their negative charges. Similarly, a larger value of chemical potential for the $d$ and $s$ quarks would result in a greater number of negative charges (particles) over the number of positive charges (antiparticles), which would increase the negative value of the Seebeck coefficient. Regardless of the species of quark, we have noticed a significant decrease in the magnitude of the Seebeck coefficient in the novel RTA model as compared to the standard RTA model. The disparity in electric charges between $u$ and $d$ quarks is apparent in the substantial discrepancy in the Seebeck coefficient, with the Seebeck coefficient of the $d$ quark being around twice that of the $u$ quark. However, it is worth noting that the electric charge of the $s$ quark is identical to that of the $d$ quark. Thus, they show a significant similarity in the values of their respective individual Seebeck coefficients. Figure \ref{5f2} illustrates the changes in the total Seebeck coefficient ($S$) in both the standard RTA and the novel RTA models as the temperature varies at different finite chemical potentials. A significant reduction in the quantity of $S$ is seen in case of the novel RTA model. In the novel RTA model, the collision integral has been modified to more accurately respect the microscopic conservation laws and the energy-dependent relaxation times. This results in significant distinctions between the RTA model and the novel RTA model. In contrast to the standard RTA model, which presumes a constant relaxation time, the novel RTA model integrates an energy-dependent scattering rate, resulting in a suppression of the transport coefficients. The energy-dependent relaxation time in the novel RTA model weakens the response of quarks to the temperature gradients. Further, the dependence of the relaxation time on energy in the novel RTA model results in the systematic reduction of the values of $Q_1$ and $Q_2$ as compared to their counterparts in the standard RTA model. Consequently, the Seebeck coefficient gets reduced 
in the novel RTA model as compared to the standard RTA model. 

Figure \ref{5f2} also demonstrates that the rate of decline of Seebeck coefficient gets gradually smaller at higher temperatures as compared to that at lower temperatures. Our findings indicate that the magnitude of $S$ falls as the temperature enhances and increases when we increase the chemical potential while keeping the temperature fixed, comparable to the behavior seen in individual quark cases. One notable observation when comparing to the standard RTA model is that, despite both $S_d$ and $S_s$ being negative, the magnitudes of $S_u$, $S_d$ and $S_s$ are such that the overall Seebeck coefficient of the medium is positive in the low temperature region and becomes slightly negative in the high temperature region at temperatures 0.34 GeV, 0.42 GeV, and 0.46 GeV for chemical potentials 0.04 GeV, 0.06 GeV, and 0.08 GeV, respectively. This transition point is crucial as it indicates a shift in the dominating charge carrier contribution within the medium. In our investigation, the Seebeck coefficient of QGP exhibits saturation at small negative values owing to the dominance of particular charge carriers at high temperatures (0.40 GeV to 0.64 GeV). The sign of the Seebeck coefficient describes the direction of the generated field relative to the temperature gradient, often indicating a rising temperature. A positive Seebeck coefficient indicates that the induced field aligns 
with the temperature gradient in the system. 

For a conducting medium, the presence of the thermoelectric effects will have an effect on both the electric current and the heat current. A finite value of the Seebeck coefficient can affect the electric current as well as the heat current in the medium. For instance, the electric current will be changed to $\textbf{J} = \sigma_{el}\textbf{E}-S \sigma_{el} \vec{\nabla} T(\textbf{r})$, while the thermal conductivity can be modified to $\kappa =  \kappa_{0} - T \sigma_{el}S^2 $. While the electrical and the thermal conductivities are always positive\footnote{The manifestation of positivity of the electrical conductivity can be 
observed through the generation of entropy, which is governed by the second law of thermodynamics. In the presence of an electromagnetic field, the electrical conductivity and the thermal conductivity are positive by requiring that $ T \partial_{\mu}\textbf{s}^{\mu} \geq 0 $ for the entropy current value \cite{Huang:2009ue, GAVIN1985826}.}, the Seebeck coefficient can be negative for the temperature and chemical potential ranges considered in this study (figure \ref{5f2}). Thus, a negative Seebeck coefficient results in an increase 
in the net electric current within the medium. Similarly, the value of the Seebeck coefficient 
decreases the thermal conductivity. It would be intriguing to consider the thermoelectric effects while calculating entropy production in the field of viscous hydrodynamics and viscous magnetohydrodynamics \cite{Huang:2009ue, GAVIN1985826}. 

\begin{figure}[H]
\begin{subfigure}[h]{0.5\textwidth}
\includegraphics[width=\textwidth]{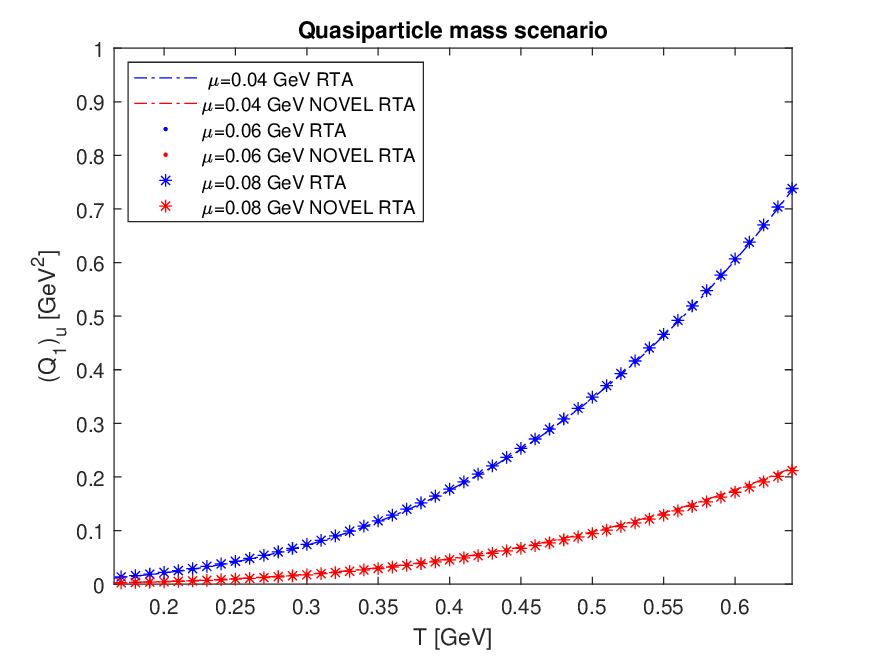}
\caption{}\label{6f1}
\end{subfigure}
\hfill
\begin{subfigure}[h]{0.5\textwidth}
\includegraphics[width=\textwidth]{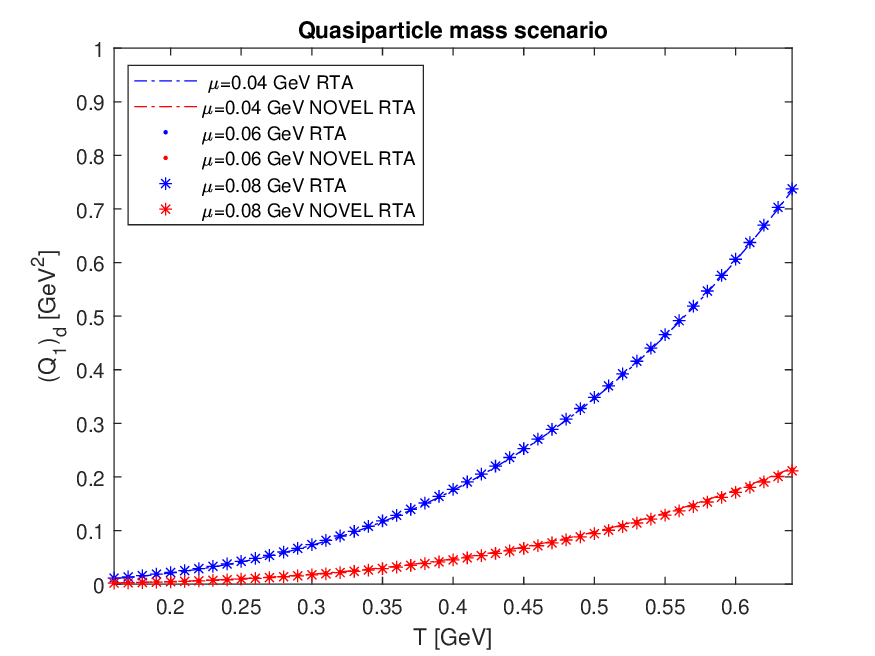}
\caption{}\label{6f2}
\end{subfigure}
\caption{The variation of $Q_1$ as a function of temperature (a) for $u$ quark and (b) for $d$ quark.}
\end{figure}

\begin{figure}[H]
\begin{subfigure}[h]{0.5\textwidth}
\includegraphics[width=\textwidth]{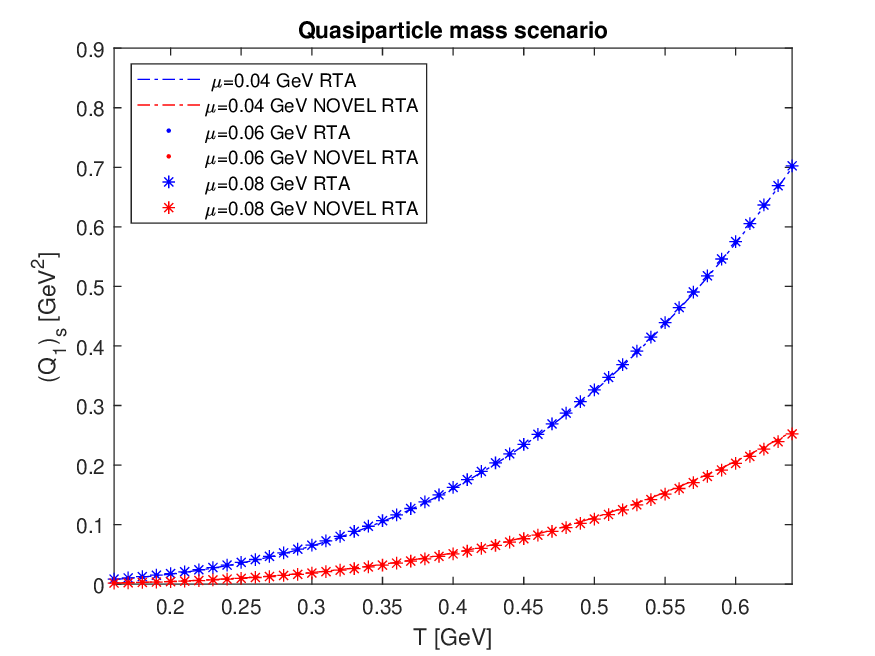}
\caption{}\label{7f1}
\end{subfigure}
\hfill
\begin{subfigure}[h]{0.5\textwidth}
\includegraphics[width=\textwidth]{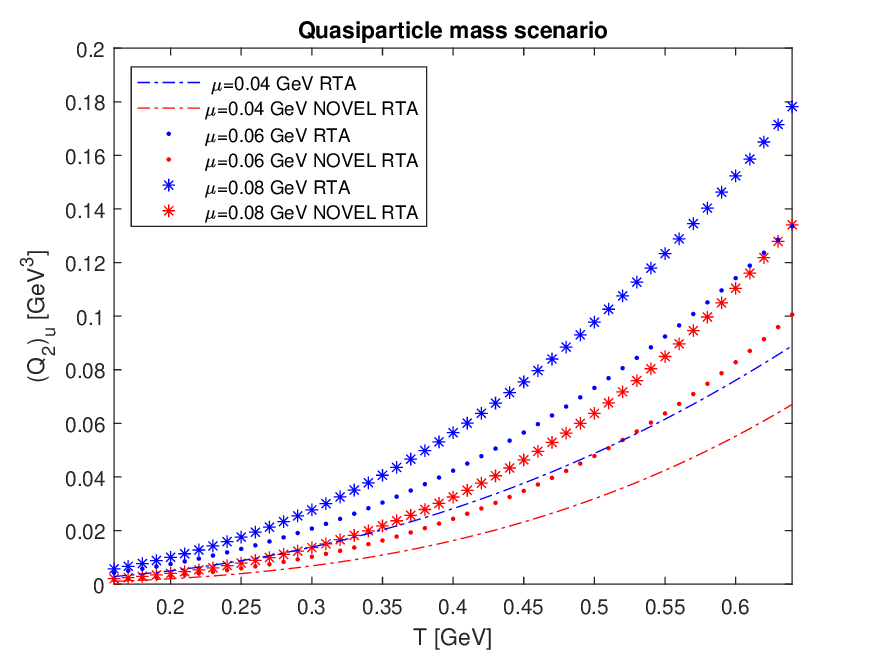}
\caption{}\label{7f2}
\end{subfigure}
\caption{(a) The variation of $Q_1$ as a function of temperature for $s$ quark and (b) the variation of $Q_2$ as a function of temperature for $u$ quark.}
\end{figure}

\begin{figure}[H]
\begin{subfigure}[h]{0.5\textwidth}
\includegraphics[width=\textwidth]{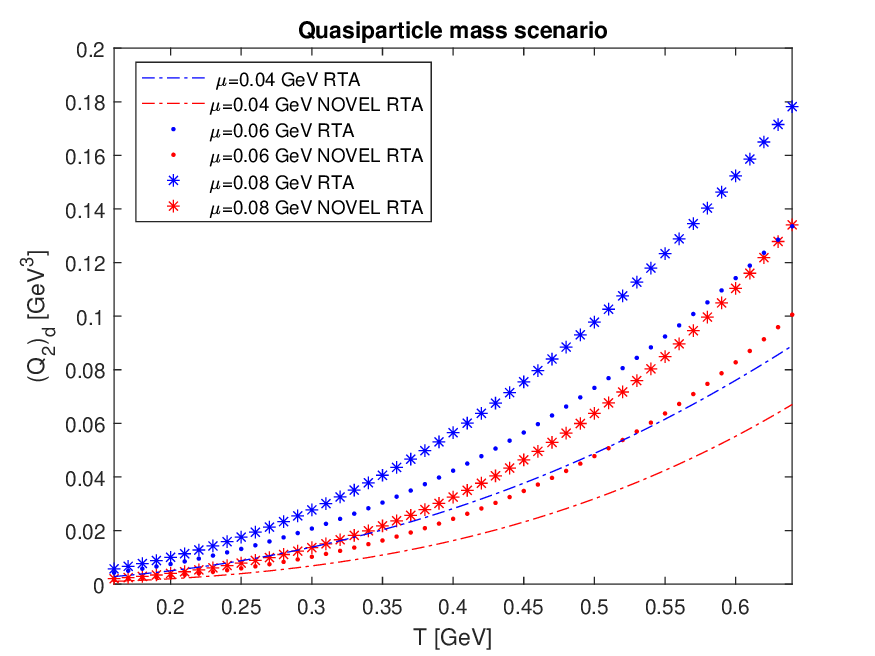}
\caption{}\label{8f1}
\end{subfigure}
\hfill
\begin{subfigure}[h]{0.5\textwidth}
\includegraphics[width=\textwidth]{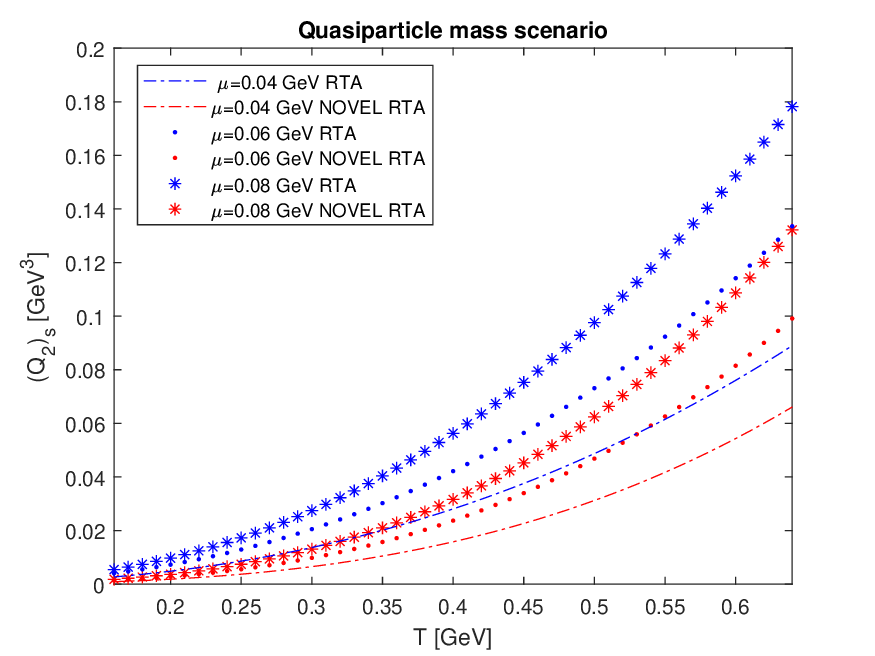}
\caption{}\label{8f2}
\end{subfigure}
\caption{The variation of $Q_2$ as a function of temperature (a) for $d$ quark and (b) for $s$ quark.}
\end{figure}

\begin{figure}[H]
\begin{subfigure}[h]{0.5\textwidth}
\includegraphics[width=\textwidth]{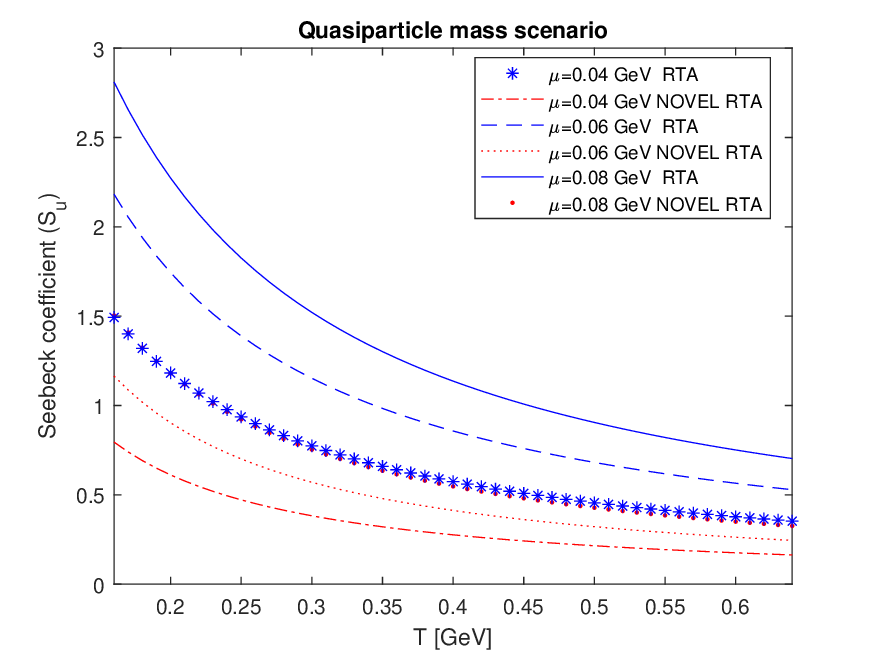}
\caption{}\label{9f1}
\end{subfigure}
\hfill
\begin{subfigure}[h]{0.5\textwidth}
\includegraphics[width=\textwidth]{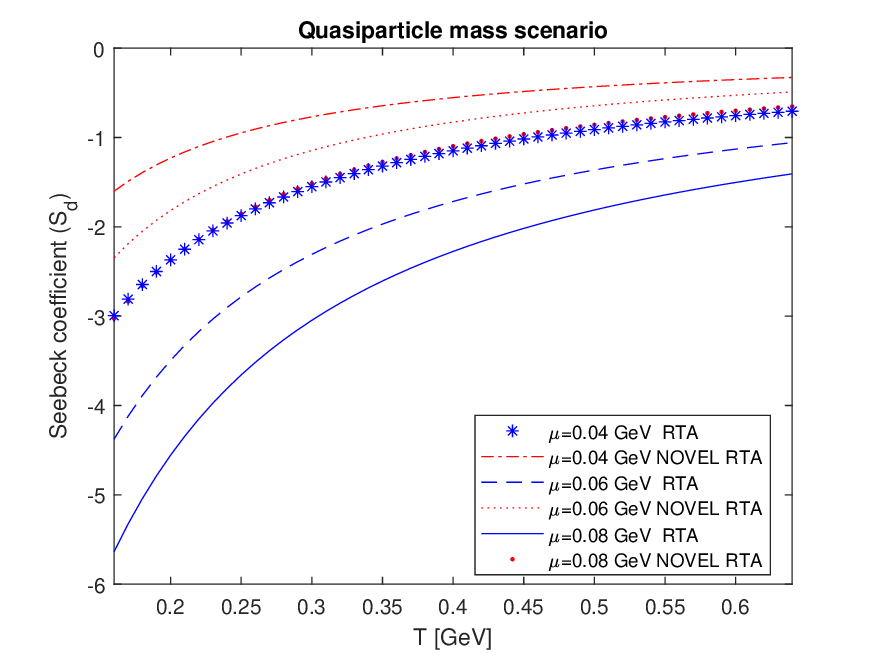}
\caption{}\label{9f2}
\end{subfigure}
\caption{The variation of Seebeck coefficient as a function of temperature (a) for $u$ quark and (b) for $d$ quark.}
\end{figure}

\begin{figure}[H]
\begin{subfigure}[h]{0.5\textwidth}
\includegraphics[width=\textwidth]{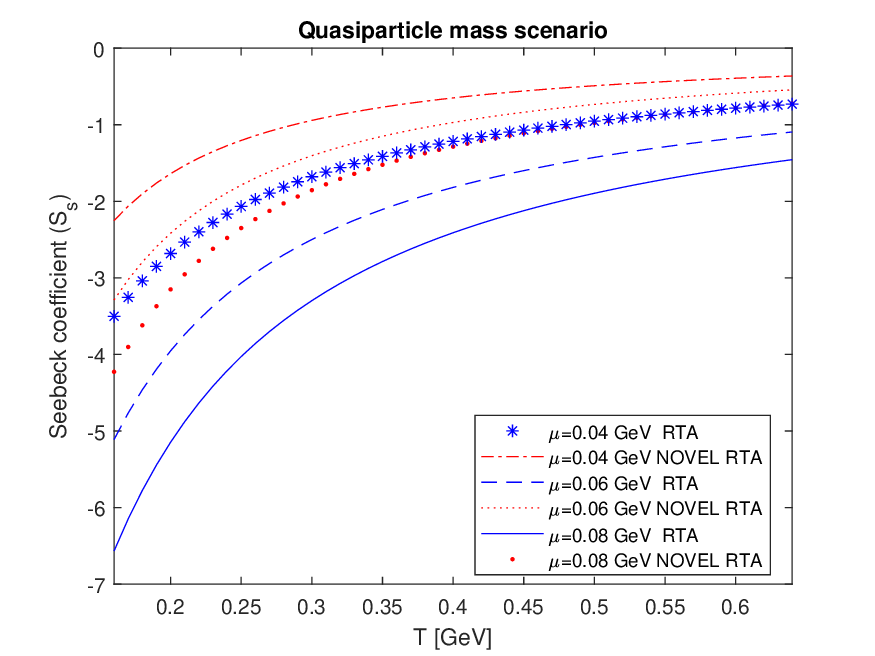}
\caption{}\label{10f1}
\end{subfigure}
\hfill
\begin{subfigure}[h]{0.5\textwidth}
\includegraphics[width=\textwidth]{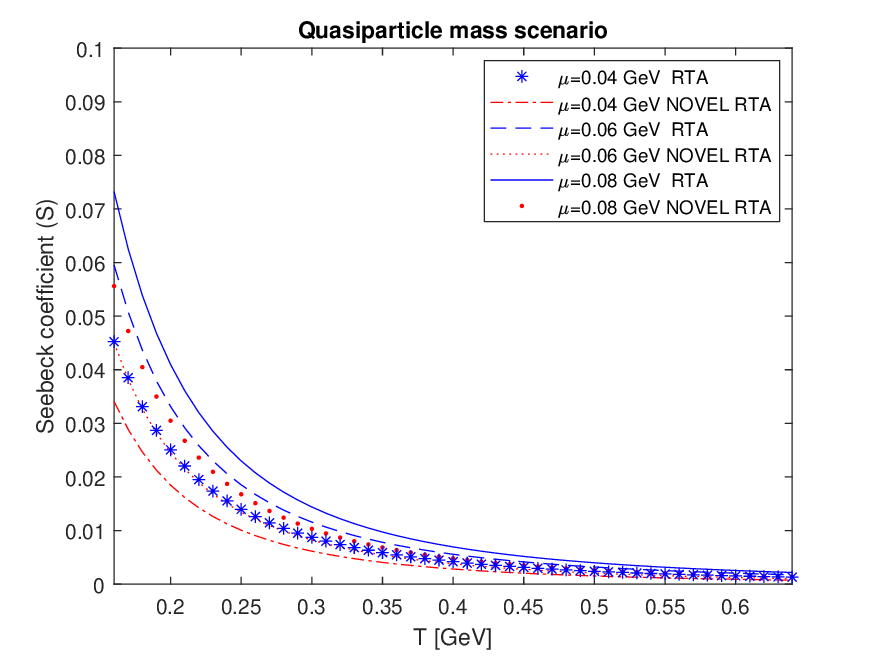}
\caption{}\label{10f2}
\end{subfigure}
\caption{(a) The variation of Seebeck coefficient as a function of temperature for $s$ quark and (b) the variation of the total Seebeck coefficient as a function of temperature.}
\end{figure}

Figures \ref{6f1}-\ref{7f1} and \ref{7f2}-\ref{8f2} respectively illustrate the variations of integrals $Q_1$ and $Q_2$ with temperature for different quark flavors ($u$, $d$ and $s$) at various finite chemical potentials using the quasiparticle framework. They have a trend that is similar to the current quark mass scenario, suggesting that they have positive values over the entire temperature range. According to the data shown in figures \ref{9f1}-\ref{10f1}, it is found that the magnitude of the Seebeck coefficient for each quark decreases as the temperature increases and increases when the chemical potential increases. The introduction of the quasiparticle description leads to a small increase in the magnitudes of the Seebeck coefficients for all the quarks. Moreover, the total Seebeck coefficient of the QGP medium in the quasiparticle description, as seen in figure \ref{10f2}, has been found to possess a slight positive value that decreases as the temperature increases and increases with the chemical potential throughout the entire temperature range. This result is different from the current mass scenario (\ref{5f2}), where the Seebeck coefficient in the high temperature region exhibited a slight negative value. The increase in magnitude of the Seebeck coefficient is the sole difference between the current quark mass scenario and the quasiparticle mass scenario. The increase of the Seebeck coefficient in the quasiparticle mass scenario indicates that the effective thermal masses of quarks influence the charge transport in the medium. In the current quark mass scenario, quarks maintain their intrinsic masses, which are small (e.g., \(m_u \simeq 3\) MeV, \(m_d \simeq 5\) MeV, \(m_s \simeq 100\) MeV). In the quasiparticle mass case, each quark gets an effective mass (eq. \eqref{28}), that is dependent upon both temperature and chemical potential through the thermal mass (eq. \eqref{29}). The temperature and chemical potential dependences of the quasiparticle masses for $u$, $d$ and $s$ quarks are shown in figures \ref{u}, \ref{d} and \ref{s}, respectively. Using the hard thermal loop perturbation theory, the thermal mass is computed and this mass modifies the energy dispersion relation. The introduction of an effective mass alters the transport properties of quarks and impacts the Seebeck coefficient. Since the effective mass modifies the quark distribution function, it shifts the phase space occupation and increases the contribution of certain momentum modes, leading to an enhanced Seebeck coefficient. Further, the presence of an effective quark mass suppresses the contribution of high-momentum quark states. This effect leads to a stronger charge separation in response to a temperature gradient, thereby increasing the Seebeck coefficient of the QGP medium. 

\section{Summary}
This work focused on the novel RTA model to investigate the thermoelectric characteristics of a dense QGP medium by determining the Seebeck coefficient. We applied the kinetic theory method to solve the 
relativistic Boltzmann transport equation using a modified collision integral for precisely determining the Seebeck coefficient of the stated medium. This study examined both the current quark mass and 
quasiparticle mass contributions to the thermoelectric effect, where both the temperature and the chemical potential affect the individual component masses in the medium. The Seebeck coefficients were computed separately for the up, down and strange quarks. These distinct coefficients were then used to determine the Seebeck coefficient of the medium by a weighted average. The magnitude of the Seebeck coefficient decreases as the temperature of the medium increases and increases with increasing value of the chemical potential. The magnitude of the Seebeck coefficient was found to be increased for the quasiparticle mass scenario as 
compared to the current quark mass scenario. 

\section{Acknowledgments}
One of the present authors (S. R.) acknowledges financial support from ANID Fondecyt Postdoctoral Grant 3240349 and S. D. acknowledges the SERB Power Fellowship, SPF/2022/000014 for the support on this work. 

\begin{appendix}
\renewcommand{\theequation}
{A.\arabic{equation}}
\section{Derivation of equation (\ref{13})}\label{appendix A}
The collision integral in the novel RTA model is given by
\begin{multline}\label{A1}	
	\textbf{C}\bigl[\textit{f}_\textit{f}\hspace{1mm}\bigr]=-\frac{\omega_f}{\tau_{fp}}\biggr[\delta f_f-
	\frac{\left\langle({\omega_f}/{\tau_{fp}})\delta f_f\right\rangle_0}{\left\langle{\omega_f}/{\tau_{fp}}\right\rangle_0}+
	P^{(0)}_1\frac{\left\langle({\omega_f}/{\tau_{fp}})P^{(0)}_1\delta f_f\right\rangle_0}{\left\langle({\omega_f}/{\tau_{fp}})P^{(0)}_1P^{(0)}_1\right\rangle_0}
	+ p^{\left\langle\mu\right\rangle}\frac{\left\langle({\omega_f}/{\tau_{fp}})p_{\left\langle\mu\right\rangle}\delta f_f\right\rangle_0}{(1/3)\left\langle({\omega_f}/{\tau_{fp}})p_{\left\langle\mu\right\rangle}p^{\left\langle\mu\right\rangle}\right\rangle_0}\Biggl].	
\end{multline}
The second term appearing in eq. \eqref{A1} can be reduced to 
		\begin{multline}\label{A2}	
		\frac{\left\langle({\omega_f}/{\tau_{fp}})\delta f_f\right\rangle_0}{\left\langle{\omega_f}/{\tau_{fp}}\right\rangle_0}
		=({1}/{\left\langle{\omega_f}/{\tau_{fp}}\right\rangle_0})\int dP \frac{\omega_f}{\tau_{fp}}{ f_{eq,f}} \delta f_f
		=\left( {1}/{\left\langle{\omega_f}/{\tau_{fp}}\right\rangle_0}\right)\\\times \Biggr[ \delta f_f \int dP \frac{\omega_f}{\tau_{fp}}{ f_{eq,f}}- \int \Bigr[ \frac{d (\delta f_{f})}{dP} \int dP \frac{\omega_f}{\tau_{fp}}f_{eq,f}\Bigl]  dP\Biggl]\\=\left( {1}/{\left\langle{\omega_f}/{\tau_{fp}}\right\rangle_0}\right)  \delta f_f \left( {\left\langle{\omega_f}/{\tau_{fp}}\right\rangle_0}\right) 
		= \delta f_f.
		\end{multline}
Similarly, the 3rd and 4th terms appearing in eq. \eqref{A1} can be respectively reduced to 
\begin{equation}\label{A3}
	P^{(0)}_1\frac{\left\langle({\omega_f}/{\tau_{fp}})P^{(0)}_1\delta f_f\right\rangle_0}{\left\langle({\omega_f}/{\tau_{fp}})P^{(0)}_1P^{(0)}_1\right\rangle_0}=P^{(0)}_1\frac{\left\langle({\omega_f}/{\tau_{fp}})P^{(0)}_1\right\rangle_0}{\left\langle({\omega_f}/{\tau_{fp}})P^{(0)}_1P^{(0)}_1\right\rangle_0}\delta f_f,	
	\end{equation}
\begin{equation}\label{A4}
p^{\left\langle\mu\right\rangle}\frac{\left\langle({\omega_f}/{\tau_{fp}})p_{\left\langle\mu\right\rangle}\delta f_f\right\rangle_0}{(1/3)\left\langle({\omega_f}/{\tau_{fp}})p_{\left\langle\mu\right\rangle}p^{\left\langle\mu\right\rangle}\right\rangle_0}\\=p^{\left\langle\mu\right\rangle}\frac{\left\langle({\omega_f}/{\tau_{fp}})p_{\left\langle\mu\right\rangle}\right\rangle_0}{(1/3)\left\langle({\omega_f}/{\tau_{fp}})p_{\left\langle\mu\right\rangle}p^{\left\langle\mu\right\rangle}\right\rangle_0}\delta f_f	
.\end{equation}
Finally, eq. (\ref{A1}) takes the following form, 
\begin{multline}\label{A5}	
	-\frac{\omega_f}{\tau_{fp}}\biggr[\delta f_f-
	\frac{\left\langle({\omega_f}/{\tau_{fp}})\delta f_f\right\rangle_0}{\left\langle{\omega_f}/{\tau_{fp}}\right\rangle_0}+
	P^{(0)}_1\frac{\left\langle({\omega_f}/{\tau_{fp}})P^{(0)}_1\delta f_f\right\rangle_0}{\left\langle({\omega_f}/{\tau_{fp}})P^{(0)}_1P^{(0)}_1\right\rangle_0}
	+ p^{\left\langle\mu\right\rangle}\frac{\left\langle({\omega_f}/{\tau_{fp}})p_{\left\langle\mu\right\rangle}\delta f_f\right\rangle_0}{(1/3)\left\langle({\omega_f}/{\tau_{fp}})p_{\left\langle\mu\right\rangle}p^{\left\langle\mu\right\rangle}\right\rangle_0}\Biggl]=\\-\frac{\omega_f A}{\tau_{fp}} \delta f_f
,\end{multline}
where 
\begin{multline}
A=\left[1-\frac{\omega_f\int p^2 (f_{eq,f}/ \tau_{fp}) dp}{\int p^2\omega_f (f_{eq,f}/\tau_{fp})dp}\right]\frac{\int p^2 (f_{eq,f}/\tau_{fp})\Bigr[1-\Bigr(\frac{\int p^2 (f_{eq,f}/\tau_{fp})dp}{\int p^2 \omega_f (f_{eq,f}/\tau_{fp})dp}\Bigl)\omega_f\Bigl] dp}{\int p^2 (f_{eq,f}/\tau_{fp})\Bigr[1-\Bigr(\frac{\int p^2 (f_{eq,f}/\tau_{fp})dp}{\int p^2 \omega_f (f_{eq,f}/\tau_{fp})dp}\Bigl)\omega_f\Bigl]^2 dp}\\ +\frac{3p\int p^3 (f_{eq,f}/ \tau_{fp}) dp}{\int p^4 (f_{eq,f}/ \tau_{fp}) dp}
.\end{multline}

\end{appendix}

\end{document}